\documentclass[aps,prb,superscriptaddress,amsfonts,amsmath,amssymb,showpacs,floatfix,reprint,longbibliography,footinbib]{revtex4-2}%

\usepackage{graphicx}
\usepackage{amsmath,amssymb,nicefrac}
\usepackage{dcolumn}
\usepackage{bm}
\usepackage{color}
\usepackage{hyperref}
\hypersetup{colorlinks=true,breaklinks,linkcolor=blue,urlcolor=blue,citecolor=blue}
\usepackage{xspace}
\usepackage{multirow}
\usepackage{physics}
\usepackage[version=4]{mhchem}
\usepackage{ulem}

\newcommand{\cso}{\ce{CsO2}}

\begin{document}

\title{Spin-Orbital Ordering in Alkali Superoxides}

\author{Kohei Shibata}
\affiliation{Department of Physics, Okayama University, Okayama 700-8530, Japan}
\author{Makoto Naka}
\affiliation{School of Science and Engineering, Tokyo Denki University, Saitama 350-0394, Japan}
\author{Harald O. Jeschke}
\affiliation{Research Institute for Interdisciplinary Science, Okayama University, Okayama 700-8530, Japan}
\author{Junya Otsuki}
\affiliation{Research Institute for Interdisciplinary Science, Okayama University, Okayama 700-8530, Japan}

\date{\today}

\begin{abstract}
Akali superoxides \ce{$A$O2} ($A=\textrm{Na}$, K, Rb, Cs), due to an open $p$ shell of the oxygen ion \ce{O2}$^-$ with degenerate $\pi$ orbitals, have spin and orbital degrees of freedom. The complex magnetic, orbital, and structural phase transitions observed experimentally in this family of materials are only partially understood. Based on density functional theory, we derive a strong-coupling effective model for the isostructural compounds \ce{$A$O2} ($A=\textrm{K}$, Rb, Cs) from a two-orbital Hubbard model. We find that {\cso} has highly frustrated exchange interactions in the $a$-$b$ plane, while the frustration is weaker for 
\ce{RbO2} and \ce{KO2}.
We solve the resulting Kugel-Khomskii model in the mean-field approximation. We show that {\cso} exhibits an antiferro-orbital (AFO) order with the ordering vector $\bm{q}=(1,0,0)$ and a stripe antiferromagnetic order with $\bm{q}=(1/2,0,0)$, which is consistent with recent neutron scattering experiments.
We discuss the role of the $\pi$-orbital degrees of freedom for the experimentally observed magnetic transitions and interpret the as-yet-unidentified $T_\mathrm{s2}=70$\,K transition in {\cso} as an orbital ordering transition.
\end{abstract}

\maketitle

\section{Introduction}

\ce{O2} is a unique molecule that possesses a magnetic moment $S=1$ on its own. 
Solid oxygen, in which \ce{O2} molecules are aggregated by the van der Waals force, exhibits a variety of electronic properties such as antiferromagnetism, metal-to-insulator transition~\cite{Freiman2004}, and superconductivity~\cite{Shimizu1998} under temperature and pressure variations. 

Another interesting system composed of \ce{O2} molecules is an ionic crystal, where \ce{O2} molecules act as electron acceptors for the counter metal ions. Alkali superoxides  \ce{$A$O2} ($A=\textrm{Na}$, K, Rb, Cs) are famous examples of such compounds~\cite{Hesse1989}. 
The \ce{O2}$^-$ ion has three electrons in the anti-bonding $\pi_g^{\ast}$ orbitals, which consist of two orbital states with symmetries similar to $d_{zx}$ and $d_{yz}$. Hence, one hole per \ce{O2}$^-$ molecule having spin $S=1/2$ and orbital degrees of freedom dominate the low-temperature physical properties.
As a consequence, spin and orbital physics as in $d$ electron systems are expected.

Electronically, alkali superoxides exhibit insulating behavior for all temperatures.
Since the unit cell contains an odd number of electrons, the insulating behavior is ascribed to the Coulomb repulsion.
A first-principles assessment with dynamical mean-field theory for \ce{RbO2} concluded that \ce{RbO2} is indeed a Mott insulator~\cite{Kovacik2012}.
This indicates that \ce{$A$O2} ($A=\textrm{K}$, Rb, Cs) are strongly correlated electron systems consisting of $\pi$ electrons.

\begin{figure}[b]
    \centering
    \includegraphics[width=\linewidth]{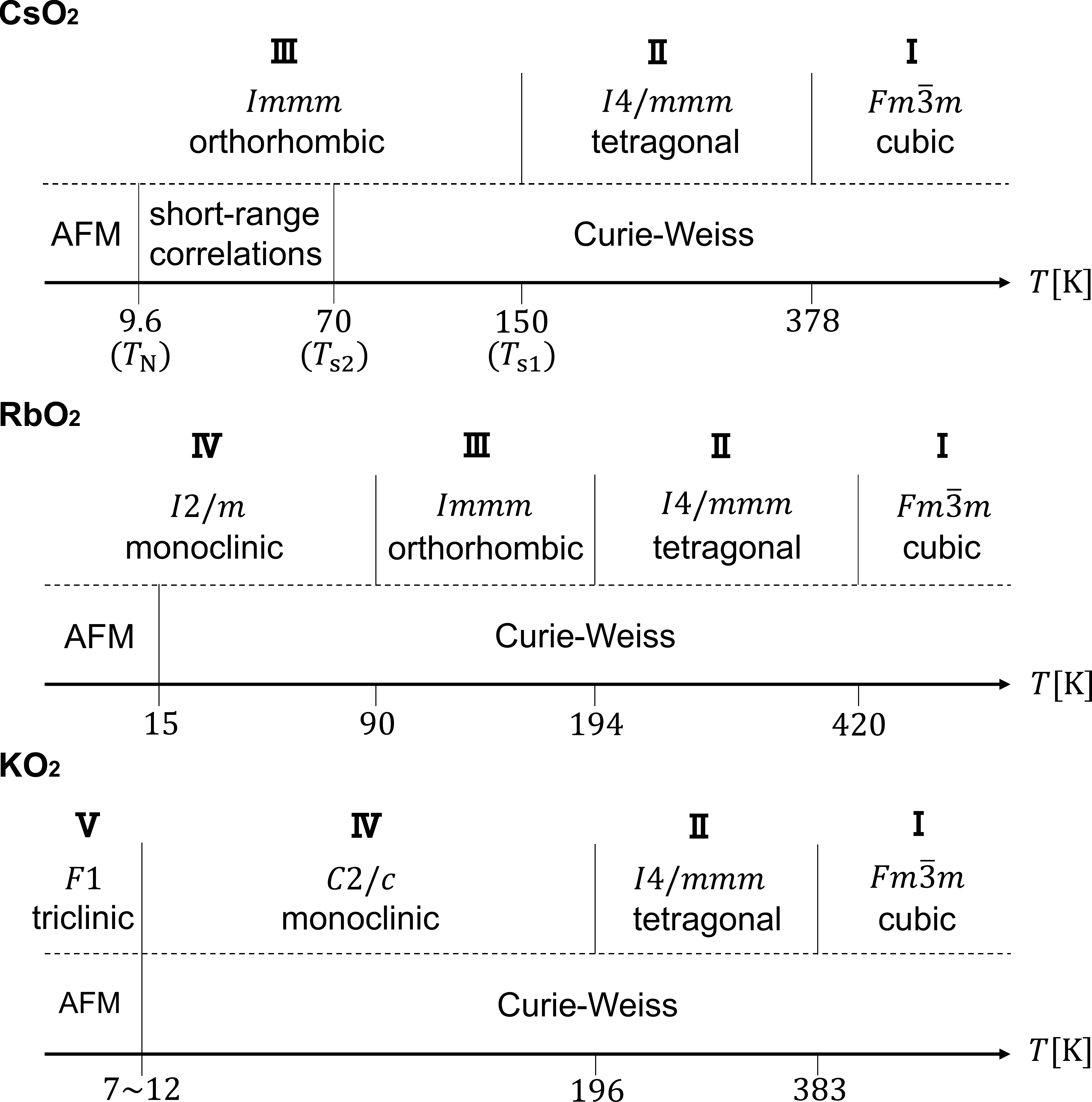}
    \caption{Summary of the structural phase transitions and magnetic properties in {\cso}, \ce{RbO2}, and \ce{KO2}. \cite{Zumsteg1974,Ziegler1976,Labhart1979,Miyajima2021}}
    \label{fig:phases}
\end{figure}

\begin{figure}[htb]
    \includegraphics[width=\linewidth]{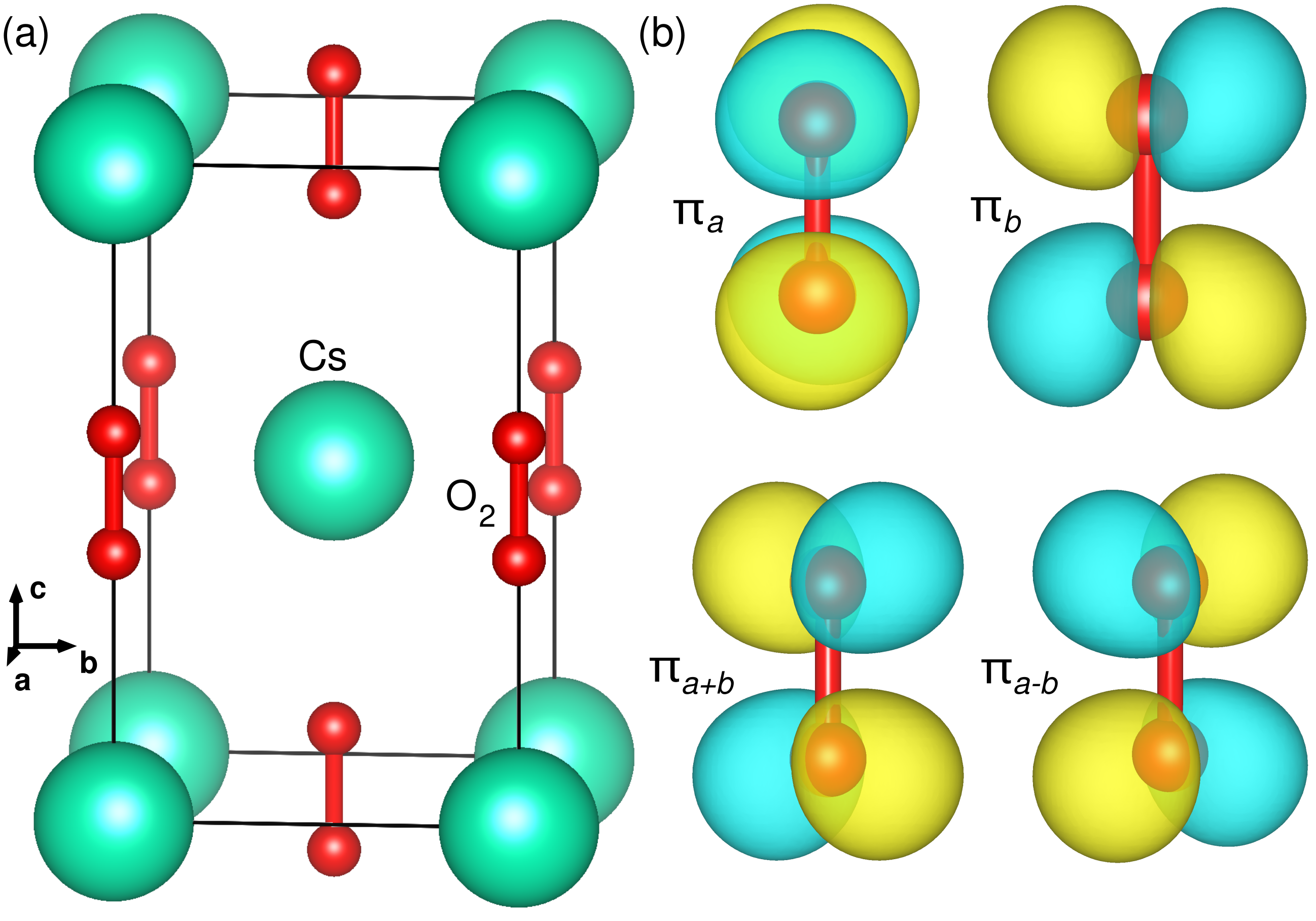}
    \caption{(a) Structure of tetragonal \ce{CsO2} ($I4/mmm$ space group). (b) Wannier functions of oxygen $\pi_g^*$ orbitals in $(\pi_a$, $\pi_b)$ basis and in $(\pi_{a+b}$, $\pi_{a-b})$ basis.
    }\label{fig:CsO2_structure}
\end{figure}

Three compounds \ce{KO2}, \ce{RbO2}, and \ce{CsO2} with the exception of \ce{NaO2} take the same crystal structure at room temperature~\cite{Zumsteg1974,Boesch1975,Ziegler1976,Labhart1979,Miyajima2021}.
The temperature variation of crystal structure and magnetic properties is summarized in Fig.~\ref{fig:phases}. 
Above 400\,K, \ce{$A$O2} exhibits the cubic NaCl-type ($Fm\bar{3}m$, no.\,225) crystal structure, in which \ce{O2} molecules are disoriented (phase I).
At around room temperature, the \ce{O2} molecules are oriented parallel to the $c$ axis (phase II), and the crystal structure of \ce{$A$O2} becomes tetragonal ($I4/mmm$, no.\,139) with $a=b<c$.
Figure~\ref{fig:CsO2_structure}\,(a) shows the crystal structure in phase II.
Phases below 200\,K are material dependent, although there is a tendency that a smaller alkali radius leads to a lower symmetry. 
\ce{KO2} undergoes two steps of symmetry lowering to monoclinic at $T=196$\,K and to triclinic at $T\simeq 10$\,K~\cite{Ziegler1976}.
\ce{RbO2} first loses the four-fold symmetry to become orthorhombic ($Immm$, no.\,71) with $a\neq b$ at $T=194$\,K and is slightly distorted to $\gamma=90.6^{\circ}$ (angle between $a$ and $b$ axes) to become monoclinic below $T=90$\,K~\cite{Miyajima2021,Astuti2019}.
\ce{CsO2} undergoes only one structural transition from tetragonal to orthorhombic ($Immm$) at $T=T_\mathrm{s1} \simeq 150$\,K~\cite{Ziegler1976,Miyajima2018}.

Magnetic properties of
\ce{KO2} and \ce{RbO2} follow the Curie-Weiss law.
A transition to the antiferromagnetic (AFM) state has been observed at $T_\mathrm{N} \simeq 10$\,K and $T_\mathrm{N} \simeq 15$\,K, respectively~\cite{Labhart1979}.
The magnetic structure of \ce{KO2} has been identified to be AFM with ordering vector $\bm{q}=(1,0,0)$ in units of the reciprocal lattice vector of the conventional unit cell~\cite{Smith1966}.
For \ce{RbO2}, the magnetic structure has not been determined, although full magnetic volume fraction has been confirmed~\cite{Astuti2019}.
On the other hand, {\cso} shows peculiar magnetic properties.
The susceptibility $\chi(T)$ in \ce{CsO2} follows the Curie-Weiss law
down to $T=T_\mathrm{s2}\simeq 70$\,K~\cite{Riyadi2012,Miyajima2018}.
Below $T_\mathrm{s2}$, $\chi(T)$ takes a maximum and is suppressed as $T$ decreases.
This indicates a development of short-range spin correlations.
It is reported that $\chi(T)$ in this region is well fitted by the Bonner-Fisher function, which was taken to suggest that the magnetic properties are described by the one-dimensional antiferromagnetic Heisenberg model~\cite{Riyadi2012,Miyajima2018}.
At $T=T_\mathrm{N}=9.6$\,K, an AFM transition takes place~\cite{Labhart1979}.
Recent neutron scattering experiments revealed a stripe-type magnetic structure~\cite{Nakano2023,Ewings2023}. Two experiments proposed different propagation vectors $\bm{q}=(0, 1/2, 0)$~\cite{Nakano2023} and $\bm{q}=(1/2, 0, 0)$~\cite{Ewings2023}, in the orthorhombic structure with $a<b$.

These experimental results demonstrate the diverse structural and magnetic properties in \ce{$A$O2} ($A=\textrm{K}$, Rb, Cs).
For a comprehensive understanding, the following two issues need to be addressed:
(i) What is the relevant microscopic control parameter that governs the physical properties in \ce{$A$O2}? 
An apparent parameter that systematically changes for different $A$ atoms is the lattice parameter, which increases in the order of K, Rb, and Cs. However, it is highly nontrivial why the short-range correlations are clearly observed only in {\cso}, which has the largest \ce{O2}--\ce{O2} distance.
We thus raise a more specific issue: (ii) What is the electronic state of {\cso}? The role of the orbitals in the magnetic properties is of particular interest.

Theoretical studies have addressed the electronic structure in  \ce{RbO2}~\cite{Kovacik2009,Ylvisaker2010,Wohlfeld2011,Kovacik2012} and \ce{KO2}~\cite{Solovyev2008,Nandy2010,Kim2010,Kim2014,Sikora2020}.
Regarding the correlated magnetic behavior in \ce{CsO2}, Riyadi \textit{et al.} proposed a zigzag orbital ordered state~\cite{Riyadi2012}.
By assuming only hopping between \ce{O2}-$\pi_g^*$ and Cs-$5p$, they argue that superexchange interaction is allowed only on a one-dimensional zigzag path in the $a$-$b$ plane.
The magnetic properties have been investigated by NMR~\cite{Klanjsek2015}, electron paramagnetic resonance~\cite{Knaflic2015}, and high-field magnetization measurement~\cite{Miyajima2018}.

In this paper, we derive an effective spin-orbital model for \ce{$A$O2} based on first-principles calculations.
We will demonstrate that geometrical frustration is the key element that constitutes a difference between $A$ atoms: The frustration plays a crucial role in {\cso} but is less important in \ce{KO2} and \ce{RbO2}.
With a mean-field calculation, we will propose an alternative type of orbital order in {\cso} that leads to the magnetic state with the experimentally observed stripe AFM order.

The rest of this paper is organized as follows.
We first derive the electronic structure of \ce{$A$O2} and an approximate tight-binding model in Sec.~II.
Using perturbation theory, we derive an effective model describing intersite spin and orbital interactions in Sec.~III.
Possible spin and orbital phase transitions are identified using the mean-field (MF) approximation in Sec.~IV.
Based on these results, we discuss implications for \ce{$A$O2} in Sec.~V.
Results are summarized in Sec.~VI.

\begin{table*}[htb]
\caption{Crystal structures used in the density functional theory calculations. O $z$ positions in brackets are obtained by structure optimization within GGA. }\label{tab:structures}
\begin{tabular}{ccc|cccc|c}
\hline
material & SG & $T$\,[K] & $a$\,[{\AA}] & $b$\,[{\AA}] & $c$\,[{\AA}] & O\,$z$ & Ref.\\\hline 
\ce{CsO2} & $Immm$ & 40 & 4.37164 & 4.40176 & 7.34214 & 0.412030 (0.408076)& \onlinecite{Miyajima2021}\\
\ce{CsO2} & $I4/mmm$ & 300 & 4.46529 & $=a$ & 7.32980 & 0.422770 (0.407922)& \onlinecite{Miyajima2021}\\\hline
\ce{RbO2} & $Immm$ & 130 & 4.14325 & 4.16334 & 7.00745 & 0.40656 (0.403463)& \onlinecite{Miyajima2021}\\
\ce{RbO2} & $I4/mmm$ & 300 & 4.20866 & $=a$ & 7.00572 & 0.407160 (0.403441)& \onlinecite{Miyajima2021}\\\hline
\ce{KO2} & $I4/mmm$ & 298 & 4.03334 & $=a$ & 6.69900 & 0.40450 (0.399067)& \onlinecite{Abrahams1955}\\\hline
\end{tabular}
\end{table*}

\section{Electronic structure}

We perform our density functional theory calculations using the all electron full potential local orbital (FPLO) basis set~\cite{Koepernik1999}. We use the generalized gradient approximation exchange correlation functional~\cite{Perdew1996}. In order to extract suitable tight-binding models,
we employ the symmetry preserving projective Wannier functions of FPLO~\cite{Koepernik2023}. We base our calculation on the structures specified in Table~\ref{tab:structures}.

\begin{figure}[t]
\includegraphics[width=0.9\linewidth]{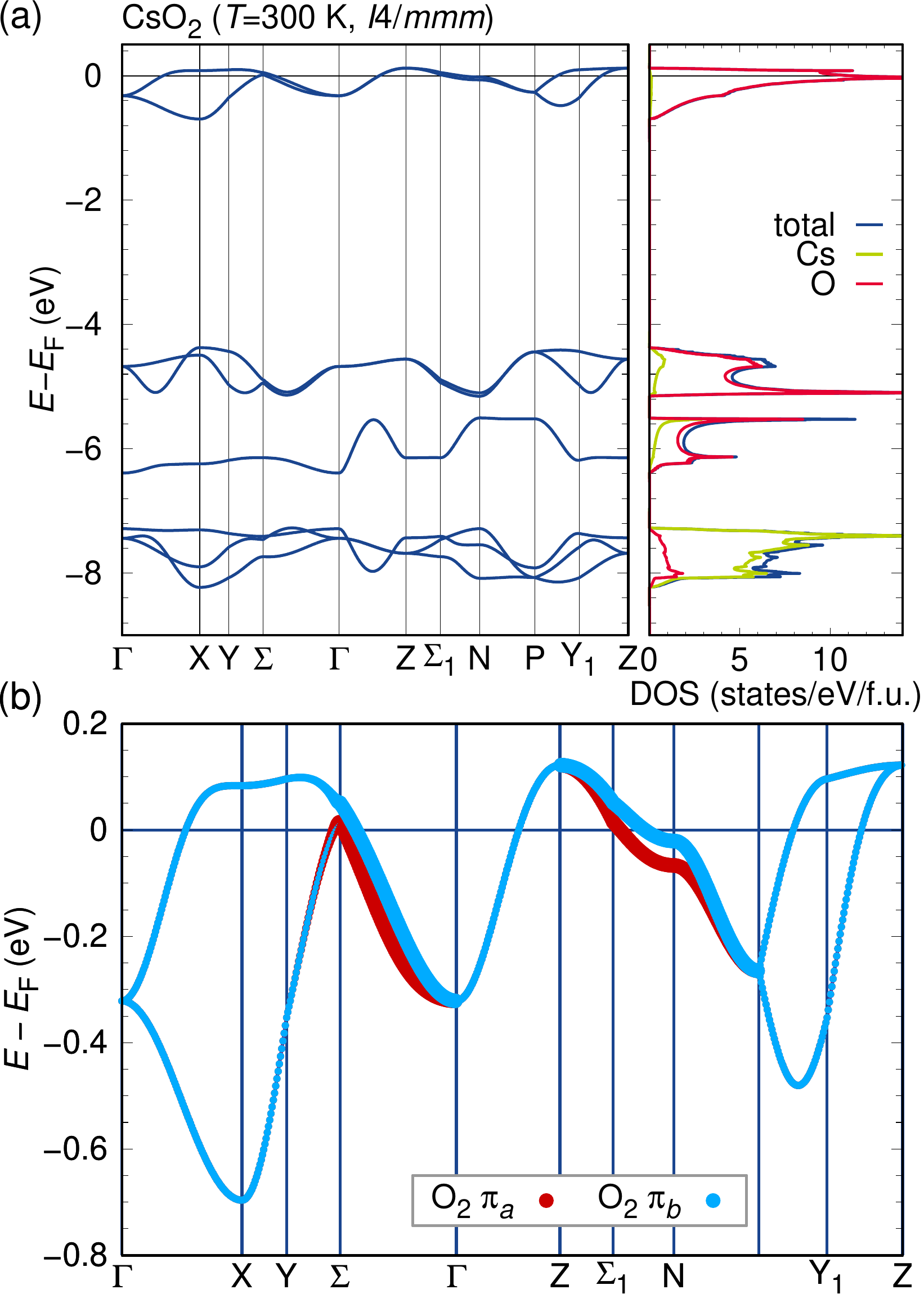}
    \caption{(a) Band structure and density of states of {\cso} in $I4/mmm$ space group ($T=300$\,K structure).
    The $\bm{k}$-points are defined in \cite{I4mmm_note}.
    (b) Wannier fit of the two
    bands near $E_{\rm F}$ with weights of the $\pi_a$ and $\pi_b$ Wannier orbitals.}
    \label{fig:DFT_CsO2_I4mmm}
\end{figure}

Figure~\ref{fig:DFT_CsO2_I4mmm}\,(a) shows the band structure and density of states of the room-temperature structure of {\cso} ($I4/mmm$ space group). The $\bm{k}$-path is a standard one for the body-centered tetragonal structure~\cite{Setyawan2010,I4mmm_note}.
There are only two $\pi_{g}^{\star}$ orbitals of oxygen near the Fermi level.
The weights of the two $\pi_g^{\ast}$ Wannier functions are shown in Fig.~\ref{fig:DFT_CsO2_I4mmm}\,(b).
The weights are not equally distributed in the path segments $Y-\Gamma$ and $Z-P$ along which either only $k_x$ or only $k_y$ changes. The dispersion of the two $\pi_{g}^{\star}$ bands of \ce{RbO2} and \ce{KO2} at room temperature ($I4/mmm$ structure) is very similar to \ce{CsO2}.

\begin{figure}[t]
\includegraphics[width=0.9\linewidth]{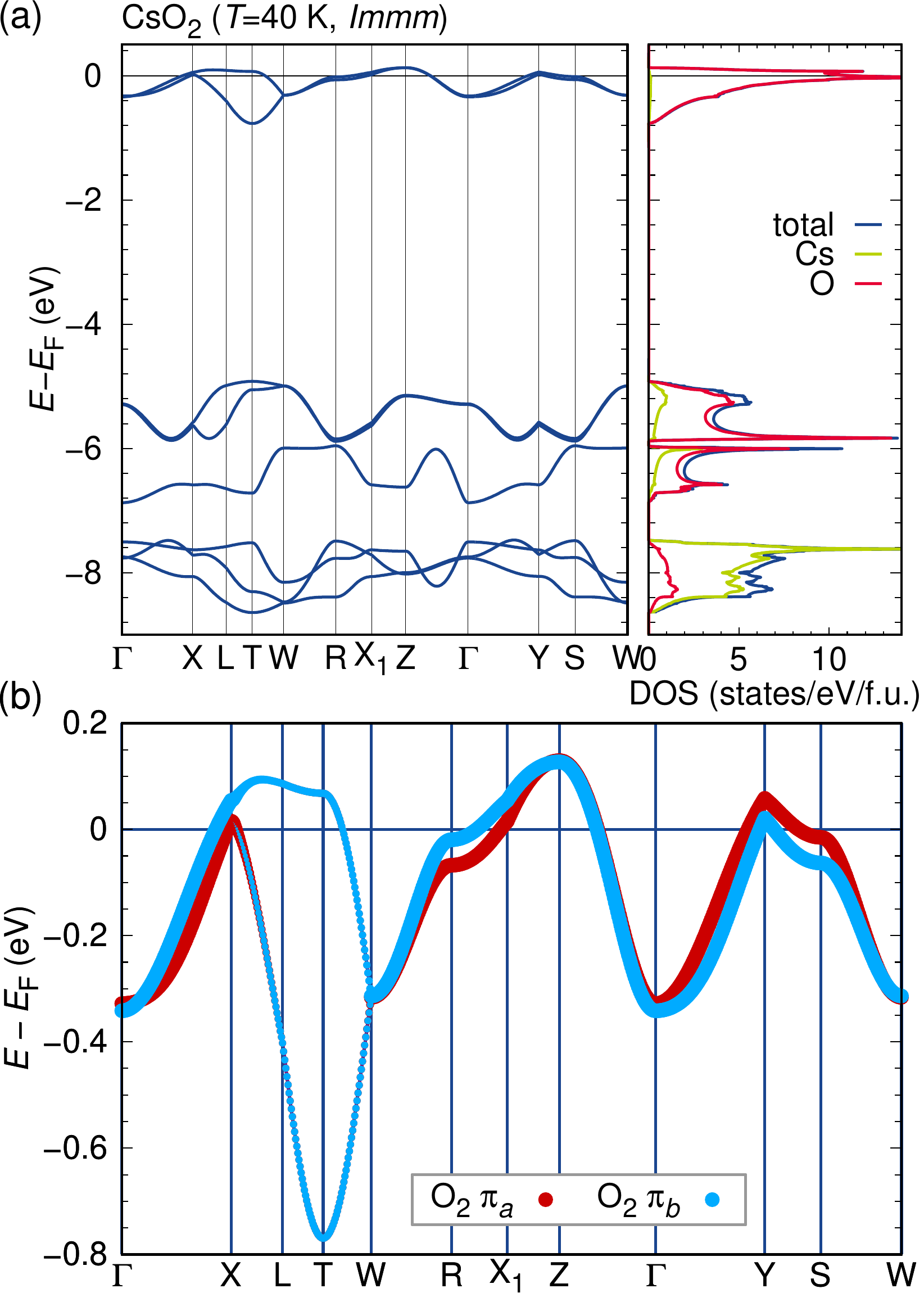}
    \caption{(a) Band structure and density of states of {\cso} in $Immm$ space group ($T=40$\,K structure).
    The $\bm{k}$-points are defined in \cite{Immm_note}.
    (b) Wannier fit of the two bands near $E_{\rm F}$ with weights of the $\pi_a$ and $\pi_b$ Wannier orbitals.}
    \label{fig:DFT_CsO2_Immm}
\end{figure}

Figure~\ref{fig:DFT_CsO2_Immm} shows the results for the $T=40$\,K structure of {\cso} ($Immm$ space group).
At $\Gamma$, the $\pi_b$ band is 14\,meV below the $\pi_a$ band because the $b$ axis is 0.7{\%} longer than the $a$ axis.
\ce{RbO2} in $Immm$ space group ($T=130$\,K structure) exhibits a similar dispersion of $\pi_{g}^{\star}$ orbitals near the Fermi level.

\begin{figure}[t]
    \centering
    \includegraphics[width=0.8\linewidth]{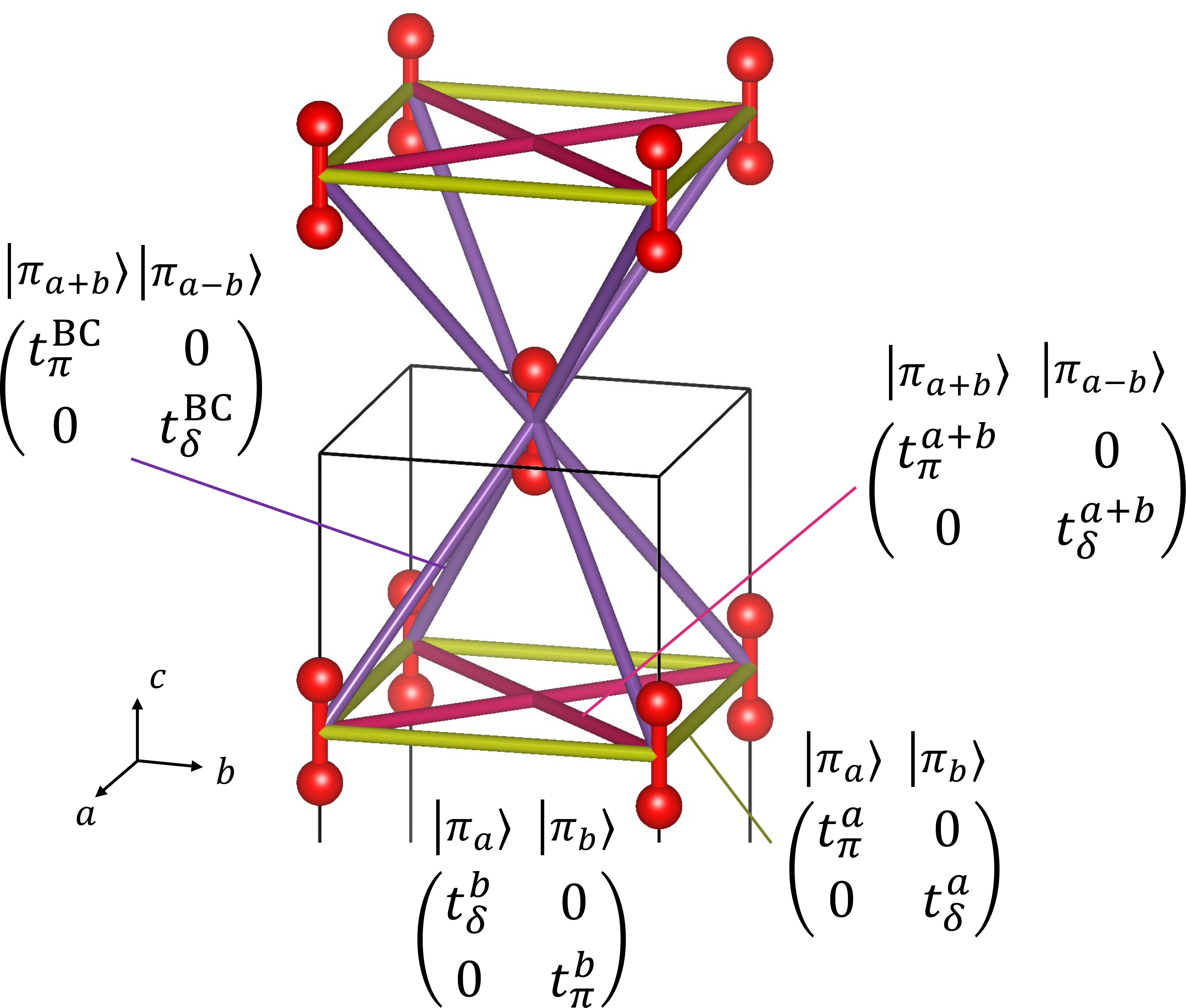}
    \caption{Relevant transfer integrals between $\pi_g^{\ast}$ orbitals on the \ce{O2^-} ions.}
    \label{fig:hopping}
\end{figure}

The band structure near the Fermi level can be well described by a two-orbital tight-binding model consisting of the $\pi_g^*$ orbitals.
Figure~\ref{fig:CsO2_structure}\,(b) shows Wannier orbitals of the $\pi_g^*$ in two sets of representations.
The ($\pi_a$, $\pi_b$) basis describes orbitals that extend along $a$ and $b$ axes, which have a symmetry similar to $d_{zx}$ and $d_{yz}$. On the other hand, ($\pi_{a+b}$, $\pi_{a-b}$) basis describes orbitals that extend to the $[110]$ and $[1\bar{1}0]$ directions.
These representations are converted to each other by
\begin{align}
    \begin{pmatrix}
    \ket{\pi_{a-b}} \\
    \ket{\pi_{a+b}}
    \end{pmatrix}
    =
    \frac{1}{\sqrt{2}}
    \begin{pmatrix}
    1 & -1 \\
    1 & 1
    \end{pmatrix}    
    \begin{pmatrix}
    \ket{\pi_{a}} \\
    \ket{\pi_{b}}
    \end{pmatrix}.
    \label{eq:orbital-rotation}
\end{align}
Figure~\ref{fig:hopping} shows three kinds of dominant bonds, which include the hopping along the $a$ or $b$ axis
(denoted by $l=a$, $b$), the diagonal hopping in the $a$-$b$ plane ($l=a+b$), and the hopping between the corner site and the body-center site ($l=\textrm{BC}$).
The hopping matrix $t_{ij}^{\gamma\gamma'}$ becomes diagonal in the ($\pi_a$, $\pi_b$) basis for $l=a$ and $b$, and diagonal in the ($\pi_{a+b}$, $\pi_{a-b}$) basis for $l=a+b$ and $\textrm{BC}$.
We assign two eigenvalues to $\pi$ and $\delta$ hopping following the usual convention~\cite{Slater1954}, and represent them by $t_{\pi}^l$ and $t_{\delta}^l$, respectively.
The hopping parameters computed using DFT and projective Wannier functions are summarized in Table~\ref{tab:hopping}.
The tight binding bands computed only with the hoppings along three paths in the tetragonal case and along four paths in the orthorhombic case perfectly reproduce the original dispersion in Figs.~\ref{fig:DFT_CsO2_I4mmm} and \ref{fig:DFT_CsO2_Immm}.

\begin{table*}[htb]
    \centering
    \caption{The hopping parameters, $t_{\pi}^l$ and $t_{\delta}^l$, for \ce{$A$O2} in units of meV. Other quantities, $t$, $\theta$, $\phi$, and $r^l$, are defined from $t_{\pi}^l$ and $t_{\delta}^l$. See Eqs.~(\ref{eq:theta_phi-1})--(\ref{eq:theta_phi-3}) for the definitions.
    ``opt O $z$'' means the use of the optimized O $z$ positions (the values in the brackets in Table~\ref{tab:structures}).}
    \label{tab:hopping}
    \begin{tabular}{cccc|rrrrrrrr|rrrrrrr}
        \hline
        & material & SG && $t^{a}_{\pi}$ & $t^{a}_{\delta}$ & $t^{b}_{\pi}$ & $t^{b}_{\delta}$ & $t^{a+b}_{\pi}$ & $t^{a+b}_{\delta}$ & $t^\mathrm{BC}_{\pi}$ & $t^\mathrm{BC}_{\delta}$ & $t$ & $\theta$ & $\phi$ & $r^{ab}$ & $r^{a+b}$ & $r^\mathrm{BC}$\\
        \hline
        (i) & \ce{CsO2} & $Immm$ && 39 & 23 & 36 & 24 & $-38$ & $-2$ & $-82$ & 22 & 84 & $16.6^{\circ}$ & $45.8^{\circ}$ & 0.63 & 0.05 & $-0.27$\\
        (ii) & \ce{CsO2} & $Immm$ & (opt O $z$) & 36 & 22 & 33 & 23 & $-34$ & 0 & $-80$ & 21 & 81 & $14.5^{\circ}$ & $43.6^{\circ}$ & 0.66 & 0.01 & $-0.26$\\
        (iii) & \ce{CsO2} & $I4/mmm$ && 55 & 27 & 55 & 27 & $-43$ & $-1$ & $-102$ & 30 & 105 & $18.8^{\circ}$ & $31.4^{\circ}$ & 0.49 & 0.02 & $-0.29$\\
        (iv) & \ce{CsO2} & $I4/mmm$ & (opt O $z$) & 32 & 19 & 32 & 19 & $-28$ & 0 & $-78$ & 20 & 79 & $12.0^{\circ}$ & $37.4^{\circ}$ & 0.59 & 0 & $-0.26$\\\hline
        (v) & \ce{RbO2} & $Immm$ && 55 & 15 & 53 & 15 & $-28$ & 0 & $-94$ & 26 & 97 & $18.9^{\circ}$ & $15.0^{\circ}$ & 0.27 & 0 & $-0.29$\\
        (vi) & \ce{RbO2} & $Immm$ & (opt O $z$) & 52 & 14 & 50 & 14 & $-26$ & 0 & $-92$ & 25 & 95 & $17.7^{\circ}$ & $14.6^{\circ}$ & 0.28 & 0.02 & $-0.27$\\
        (vii) & \ce{RbO2} & $I4/mmm$ && 52 & 13 & 52 & 13 & $-25$ & $-1$ & $-92$ & 26 & 94 & $18.2^{\circ}$ & $13.0^{\circ}$ & 0.25 & 0.04 & $-0.28$\\
        (viii) & \ce{RbO2} & $I4/mmm$ & (opt O $z$) & 48 & 12 & 48 & 12 & $-23$ & 0 & $-90$ & 25 & 92 & $16.6^{\circ}$ & $12.8^{\circ}$ & 0.25 & 0.02 & $-0.27$\\\hline
        & \ce{KO2} & $I4/mmm$ && 68 & 6 & 68 & 6 & $-19$ & $-1$ & $-109$ & 31 & 113 & $21.3^{\circ}$ & $4.5^{\circ}$ & 0.09 & 0.05 & $-0.28$\\
        & \ce{KO2} & $I4/mmm$ & (opt O $z$) & 63 & 5 & 63 & 5 & $-17$ & 0 & $-107$ & 30 & 110 & $19.3^{\circ}$ & $4.1^{\circ}$ & 0.08 & 0.02 & $-0.23$\\
        \hline
    \end{tabular}
\end{table*}

\begin{figure}[t]
    \centering
    \includegraphics[width=0.9\linewidth]{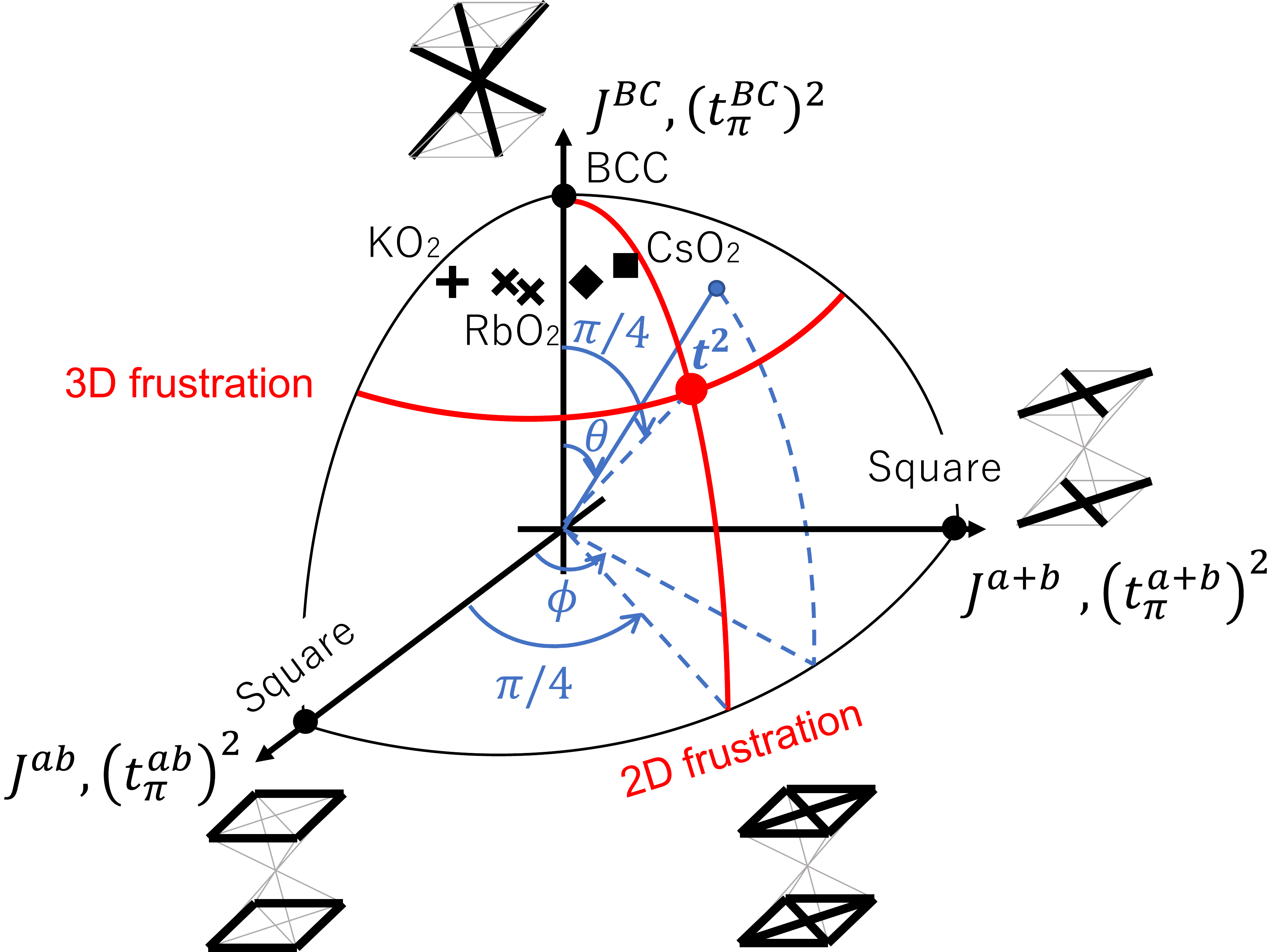}
    \caption{Parameterization of the bond dependence of the transfer integrals using polar coordinates. The symbols indicate the DFT estimates for {\cso}, \ce{RbO2}, and \ce{KO2} (only results without the optimization are shown).}
    \label{fig:polar}
\end{figure}

In order to characterize differences in the hopping parameters between the three compounds, we introduce two kinds of dimensionless parameters as follows.
Firstly, the ratio between $t_{\pi}^l$ and $t_{\delta}^l$ for each bond $l$ is defined as $r^l \equiv t_{\delta}^l/t_{\pi}^l$.
Table~\ref{tab:hopping} indicates that $|r^l|$ ranges from 0 to 0.68 depending on the bond and $A$ atoms.
Here, we averaged $r^a$ and $r^b$ as $r^{ab} \equiv (r^a+r^b)/2$.
Secondly, the relation between different bonds is represented by three-dimensional polar coordinates defined by
\begin{align}
(t_{\pi}^\textrm{BC})^2 &= t^2 \cos\theta,
\label{eq:theta_phi-1}
\\
(t_{\pi}^\textrm{ab})^2 &= t^2 \sin\theta \cos\phi,
\\
(t_{\pi}^{\textrm{a}+\textrm{b}})^2 &= t^2 \sin\theta \sin \phi,
\label{eq:theta_phi-3}
\end{align}
where $(t_{\pi}^\textrm{ab})^2 \equiv [(t_{\pi}^\textrm{a})^2+(t_{\pi}^\textrm{b})^2]/2$.
Here, we consider the square of the hopping parameters because the effective intersite exchange interactions are proportional to $t^2$ rather than $t$ itself (Sec.~\ref{sec:eff_model}).
A graphical interpretation of $\theta$ and $\phi$ is presented in Fig.~\ref{fig:polar}.
The north pole corresponds to a system with only the nearest-neighbor hopping $t^\mathrm{BC}$, while the equator corresponds to a two-dimensional square lattice model with $t^a$ and $t^{a+b}$.
Combinations of two or more hopping parameters lead to geometrical frustration of the exchange interactions. In particular, $\phi=45^{\circ}$ leads to a two-dimensional frustration within the $a$-$b$ plane, while $\theta=45^{\circ}$ leads to a three-dimensional frustration between $l=\mathrm{BC}$ bond and in-plane bonds.

DFT estimates of $\theta$ and $\phi$ are presented in Table~\ref{tab:hopping} and marked with symbols in Fig.~\ref{fig:polar}.
The \ce{$A$O2} series is located in the $\theta < 45^{\circ}$ region, which means that the $l=\mathrm{BC}$ bond is the largest and three-dimensional hopping plays a major role.
Regarding the hopping in the $a$-$b$ plane, \ce{CsO2} is located around $\phi = 45^{\circ}$, which indicates that \ce{CsO2} is characterized by two-dimensional frustration between $l=a$ and $a+b$ bonds. This frustration is largest in \ce{CsO2} and tends to get weaker for smaller alkali ions.

\section{Strong-coupling effective model}\label{sec:eff_model}

Since \ce{$A$O2} ($A=\textrm{K}$, Rb, Cs) are Mott insulators~\cite{Kovacik2012}, we employ a strong-coupling effective model that describes intersite exchange interactions between localized spin and orbital degrees of freedom.
We first compare two kinds of exchange interactions, and then derive an explicit form of the Hamiltonian.

\subsection{Comparison between two perturbation processes for the exchange interactions}

There are two perturbation processes that give rise to intersite exchange interactions between $\pi_g^*$ electrons on the \ce{O2^-} ions.
One is the second-order process of the \ce{O2^-}--\ce{O2^-} hopping, and the other is the fourth-order superexchange process of the \ce{O2^-}--\ce{Cs^+} hopping.
The coupling constant in the second-order process, $J^{(2)}$, is estimated to be
\begin{align}
    J^{(2)} \sim \frac{t^2}{U},
\end{align}
where $U$ is the Coulomb repulsion between two electrons on the same orbital. 
On the other hand, the coupling constant in the fourth-order process through \ce{Cs^+} ions, $J^{(4)}$, is estimated to be
\begin{align}
    J^{(4)} \sim \frac{(t_{\textrm{Cs-O}_2})^4}{U \Delta_{5p}^2},
\end{align}
where $t_{\textrm{Cs-O}_2}$ is the hopping amplitude between the \ce{O2}-$\pi_g^*$ orbital and the Cs-$5p$ orbital.
We estimated $t_{\textrm{Cs-O}_2}$ by projecting the energy dispersion in Fig.~\ref{fig:DFT_CsO2_Immm} onto an eight-band tight-binding model consisting of \ce{O2}-$\pi_g^*$, $\pi_u$, $\sigma_g$, and Cs-$5p$ orbitals.
We thus concluded that $t_{\textrm{Cs-O}_2}$ is the same order as $t$ in the two-band model, namely, $t \sim t_{\textrm{Cs-O}_2}$.
Hence, the ratio between $J^{(2)}$ and $J^{(4)}$ is estimated as
\begin{align}
    \frac{J^{(4)}}{J^{(2)}} \sim \frac{t^2}{\Delta_{5p}^2} \sim 10^{-4}.
\end{align}
Here, we used $t\sim 0.1$\,eV (Table~\ref{tab:hopping}) and $\Delta_{5p} \sim 8$\,eV (Fig.~\ref{fig:DFT_CsO2_Immm}).
Therefore, we consider from now on only the second-order process from the direct \ce{O2}--\ce{O2} hopping, neglecting the fourth-order superexchange process through Cs.

\subsection{Derivation of the interactions}

In order to derive the spin-orbital exchange interactions, we begin with the two-orbital Hubbard model consisting of the $\pi_g^{*}$ orbitals, $\pi_{a}$ and $\pi_{b}$. The Hamiltonian reads
\begin{align}
    H= &\sum_{\langle ij \rangle}\sum_{\gamma \gamma' \sigma}t^{\gamma \gamma'}_{ij}(c^{\dag}_{i\gamma\sigma}c_{j\gamma'\sigma} +\textrm{H.c.})
    + U\sum_{i\gamma}n_{i\gamma\uparrow}n_{i\gamma\downarrow}
    \notag \\
    &+ U'\sum_{i\gamma>\gamma'}n_{i\gamma}n_{i\gamma'}
    +J_\mathrm{H}\sum_{i\sigma\sigma'\gamma>\gamma'} c^{\dag}_{i\gamma \sigma}c^{\dag}_{i\gamma' \sigma'} c_{i\gamma \sigma'}c_{i\gamma' \sigma}
    \notag \\
    &+J'_\mathrm{H}\sum_{i\gamma\neq\gamma'}c^{\dag}_{i\gamma\uparrow}c^{\dag}_{i\gamma\downarrow}c_{i\gamma'\downarrow}c_{i\gamma'\uparrow},
\end{align}
where
$c_{i\gamma\sigma}$ is the annihilation operator for site $i$, orbital $\gamma$, and spin $\sigma$, and
$n_{i\gamma\sigma}=c^{\dag}_{i\gamma\sigma} c_{i\gamma\sigma}$ is the number operator.
The first term represents the electron hopping between site $i$ and site $j$.
The symbol $\langle ij \rangle$ stands for the pairs of neighboring \ce{O2} sites shown in Fig.~\ref{fig:hopping}. 
The second and third terms represent the intra-orbital and inter-orbital Coulomb repulsion, respectively.
The fourth and fifth terms represent the Hund's rule coupling and the pair hopping interaction, respectively.

\begin{figure}[t]
    \centering
    \includegraphics[width=0.7\linewidth]{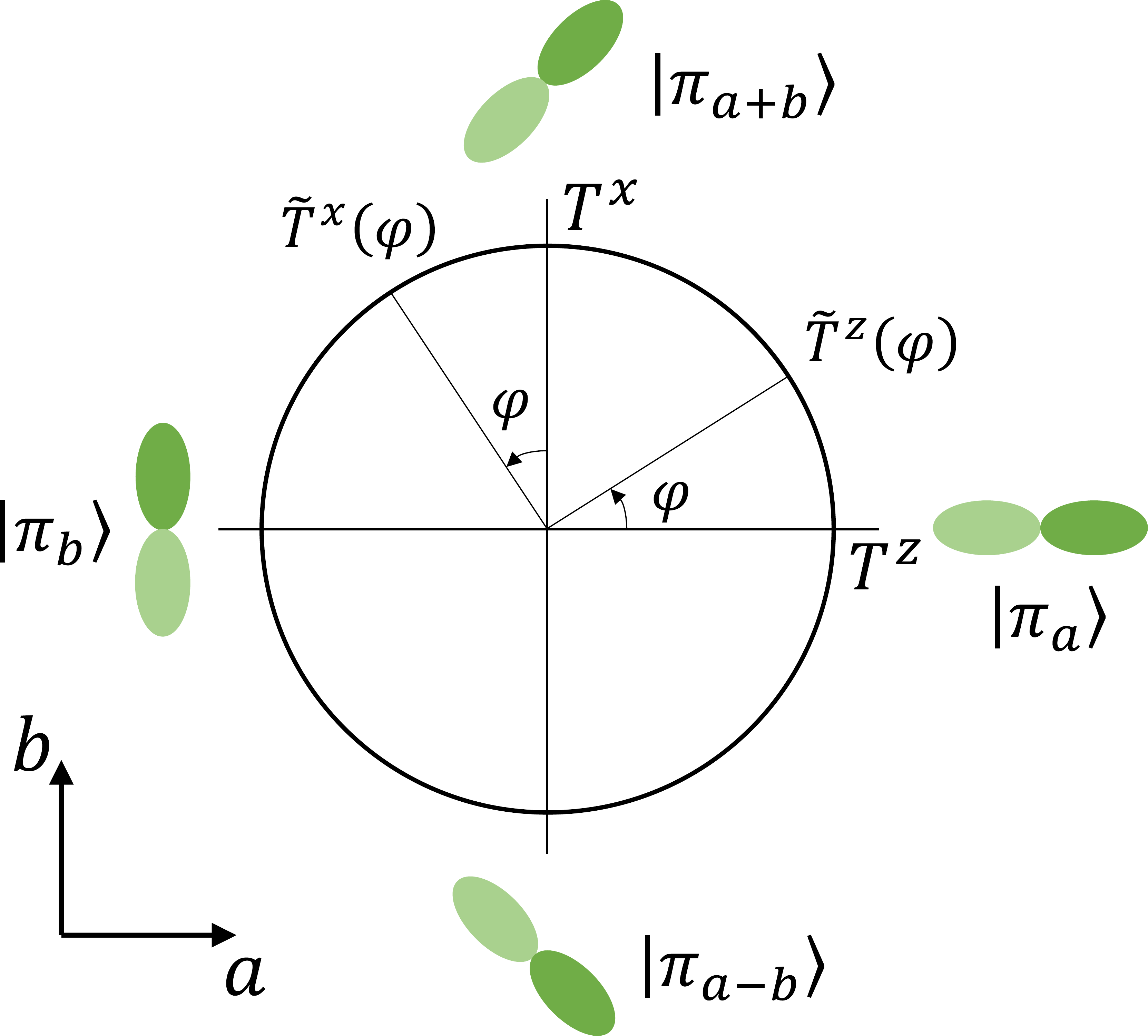}
    \caption{A diagram for the orbital (pseudospin) operators and the orbital states.
    The orbital on the right (left) represents the eigenstate of the operator $T^z$ with the eigenvalue $+1/2$ ($-1/2$). The orbital state is rotated around the $c$ axis by $\varphi/2$ as the operator is rotated by $\varphi$ in the pseudospin space.}
    \label{fig:orbital_operators}
\end{figure}

We consider the Mott insulating state with the occupation number $n=3$ per site. 
For convenience, we treat this as $n=1$ from now on, using a hole picture.
Note that the relation between orbital states and lattice distortion is reversed in the hole picture, though the effective Hamiltonian derived below is the same for $n=1$ and $n=3$.
The local electronic degrees of freedom in the Mott insulating state are described by the spin and orbital (pseudospin) operators defined by
\begin{align}
    & \bm{S}_{i} = \frac{1}{2} \sum_{\gamma \sigma \sigma'} c_{i \gamma \sigma}^{\dag} \bm{\sigma}_{\sigma \sigma'} c_{i \gamma \sigma'},
    \\
    & \bm{T}_{i} = \frac{1}{2} \sum_{\gamma \gamma' \sigma} c_{i \gamma \sigma}^{\dag} \bm{\sigma}_{\gamma \gamma'} c_{i \gamma' \sigma},
\end{align}
where $\bm{\sigma}=(\sigma^x, \sigma^y, \sigma^z)$ is the Pauli matrix.
The eigenstates of $T^z$ are $\ket{\pi_a}$ and $\ket{\pi_b}$ orbitals, which extend along the $a$ and $b$ axes, respectively [Fig.~\ref{fig:CsO2_structure}\,(b)].
On the other hand, the eigenstates of $T^x$ describe $\ket{\pi_{a+b}}$ and $\ket{\pi_{a-b}}$ orbitals that extend to the diagonal direction [Fig.~\ref{fig:CsO2_structure}\,(b)], because the ($\pi_{a+b}$, $\pi_{a-b}$) basis is obtained by linear combination of ($\pi_{a}$, $\pi_{b}$) as given in Eq.~(\ref{eq:orbital-rotation}).
In general, we can describe the orbital state in an arbitrary direction by rotating $T^z$ and $T^x$ in the pseudospin space as
\begin{align}
    \widetilde{T}^z(\varphi) &= T^z \cos \varphi + T^x \sin \varphi,
    \notag \\
    \widetilde{T}^x(\varphi) &= -T^z \sin \varphi + T^x \cos \varphi.
    \label{eq:rotate_orbital_ops}
\end{align}
Figure~\ref{fig:orbital_operators} shows the variation of the eigenstates of the operators $\widetilde{T}^z(\varphi)$ and $\widetilde{T}^x(\varphi)$. The orbital state is rotated by $\varphi/2$ around the $c$ axis.

\begin{figure}[t]
    \centering
    \includegraphics[width=0.9\linewidth]{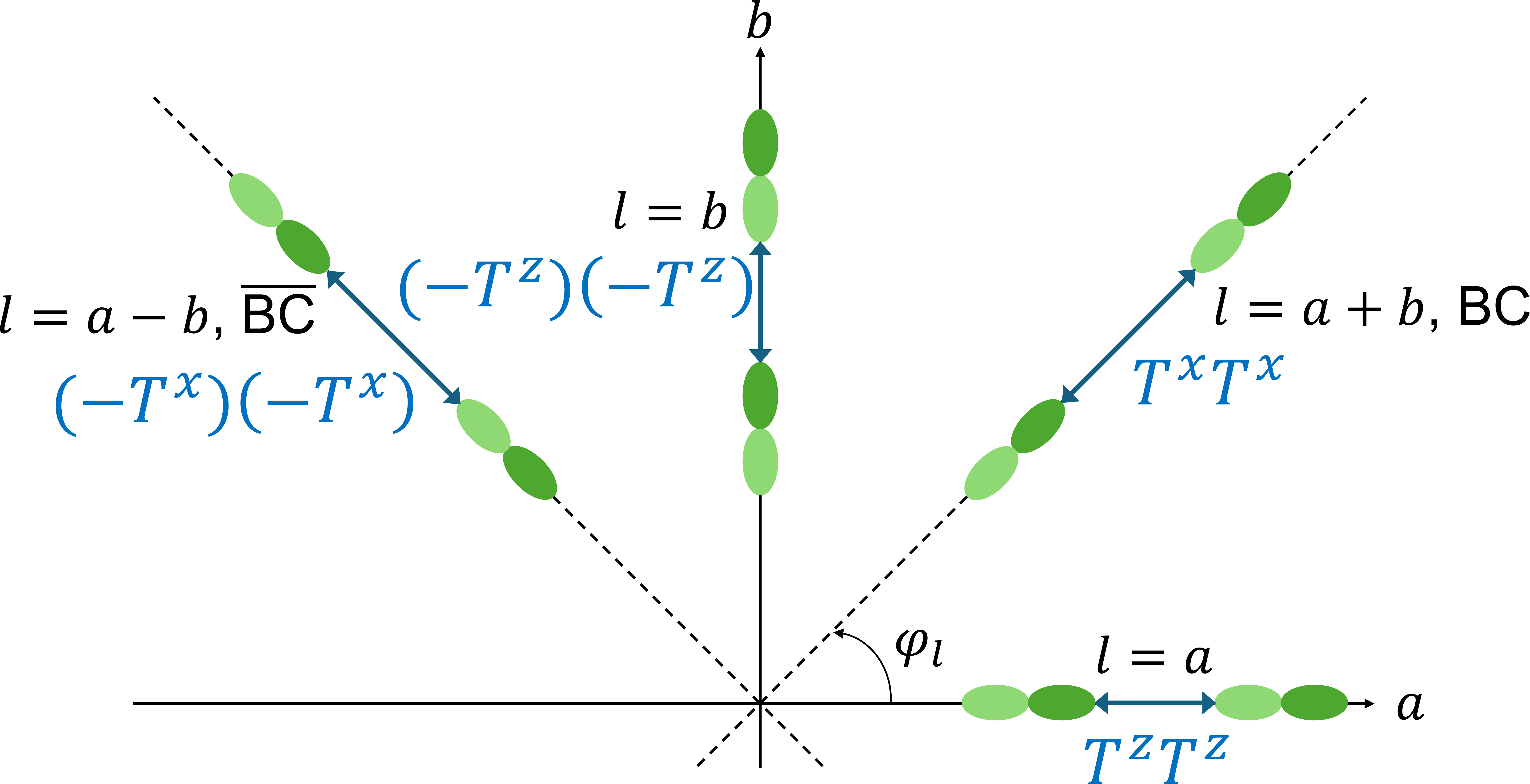}
    \caption{The orbital-orbital interactions $\tau_i^l \tau_j^l$ described by the bond-dependent orbital operator $\tau_i^l$.}
    \label{fig:interaction}
\end{figure}

In second-order perturbation theory around the atomic limit with respect to the hopping, we can derive Kugel-Khomskii (KK) interactions~\cite{Kugel1982} for alkali superoxides. We present only the result below, and refer readers to Refs.~\cite{Maekawa2004,Igoshev2023} for general derivation of the KK interaction. The effective Hamiltonian $\mathcal{H}_\mathrm{eff}$ for the $n=1$ subspace of the $\pi_g^{\ast}$ system is given by
\begin{widetext}
\begin{align}
    \mathcal{H}_\mathrm{eff}
    = &-2\sum_{\langle ij \rangle}J_{1}^{l} \left( \frac{3}{4}+\bm{S}_{i}\cdot\bm{S}_{j} \right) \left\{ \left[ 1+\left(r^{l}\right)^{2} \right] \left( \frac{1}{4}-\tau_{i}^{l}\tau_{j}^{l} \right) -r^{l}\left( \tau_{i}^{l+}\tau_{j}^{l-}+\tau_{i}^{l-}\tau_{j}^{l+} \right) \right\}
    \notag \\
    &-2\sum_{\langle ij \rangle}J_{2}^{l}\left(\frac{1}{4}-\bm{S}_{i}\cdot\bm{S}_{j}\right) \left\{\left[1+\left(r^{l}\right)^{2} \right]\left(\frac{1}{4}-\tau_{i}^{l}\tau_{j}^{l}\right) +r^{l}\left(\tau_{i}^{l+}\tau_{j}^{l-}+\tau_{i}^{l-}\tau_{j}^{l+}\right)\right\}
    \notag \\
    &-2\sum_{\langle ij \rangle}J_{2}^{'l}\left(\frac{1}{4}-\bm{S}_{i}\cdot\bm{S}_{j}\right) \left[\left(\frac{1}{2}+\tau_{i}^{l}\right)\left(\frac{1}{2}+\tau_{j}^{l}\right)+\left(r^{l}\right)^{2}\left(\frac{1}{2}-\tau_{i}^{l}\right)\left(\frac{1}{2}-\tau_{j}^{l}\right)-r^{l}\left(\tau_{i}^{l+}\tau_{j}^{l+}+\tau_{i}^{l-}\tau_{j}^{l-}\right)\right]
    \notag \\
    &-2\sum_{\langle ij \rangle}J_{3}^{l}\left(\frac{1}{4}-\bm{S}_{i}\cdot\bm{S}_{j}\right) \left[\left(\frac{1}{2}+\tau_{i}^{l}\right)\left(\frac{1}{2}+\tau_{j}^{l}\right)+\left(r^{l}\right)^{2}\left(\frac{1}{2}-\tau_{i}^{l}\right)\left(\frac{1}{2}-\tau_{j}^{l}\right)+r^{l}\left(\tau_{i}^{l+}\tau_{j}^{l+}+\tau_{i}^{l-}\tau_{j}^{l-}\right)\right],
    \label{eq:Heff}
\end{align}
\end{widetext}
where $l$ is the bond index that takes the values $a$, $b$, $a+b$, $a-b$, $\textrm{BC}$, or $\overline{\textrm{BC}}$, depending on the combination of $i$ and $j$ (Fig.~\ref{fig:interaction}).
Here, we introduced $l=a-b$ and $\overline{\textrm{BC}}$, which have the same hopping parameter as $l=a+b$ and $\textrm{BC}$, respectively, but are different in the definition of the operator $\tau_i^l$ (see below).
The coupling constants $J_{1}^{l}, J_{2}^{l}, J_{2}^{'l}$, and $J_{3}^{l}$ are given by
\begin{align}
    J_{1}^{l} &= \frac{(t_{\pi}^{l})^2}{U^{'}-J_\mathrm{H}},
    \notag \\
    J_{2}^{l} &= \frac{(t_{\pi}^{l})^2}{U^{'}+J_\mathrm{H}},
    \notag \\
    {J_{2}^{'}}^{l} &= \frac{(t_{\pi}^{l})^2}{U-J_\mathrm{H}^{'}},
    \notag \\
    J_{3}^{l} &= \frac{(t_{\pi}^{l})^2}{U+J_\mathrm{H}^{'}}.
    \label{eq:Js}
\end{align}
The operators $\tau^{l}$ and $\tau^{l\pm}$ are bond-dependent orbital operators defined by
\begin{align}
    \tau^{l} &= \widetilde{T}^z ( 2\varphi_l ),
    \notag \\
    \tau^{l\pm} &= \widetilde{T}^x ( 2\varphi_l ) \pm iT^{y},
    \label{eq:tau}
\end{align}
where $\varphi_l$ is the azimuthal angle around the $c$ axis: $\varphi_{a}=0$, $\varphi_{a+b}=\varphi_\mathrm{BC}=\pi/4$, $\varphi_{b}=\pi/2$, and $\varphi_{a-b}=\varphi_\mathrm{\overline{BC}}=3\pi/4$.
The coupling constants satisfy $J_1^l > J_2^l \simeq J_2^{'l} > J_3^l$ in a typical choice of parameters. Details are given in Appendix~\ref{sec:Js}.

Equation~(\ref{eq:Heff}) includes $\delta$ hopping, which is expressed by the coefficient $r^l \equiv t_{\delta}^l/t_{\pi}^l$.
Without $r^l$ terms, namely, inserting $r^l=0$ into Eq.~(\ref{eq:Heff}), $\mathcal{H}_\mathrm{eff}$ is reduced to the well-known form of the KK Hamiltonian for $d$-$e_g$ orbital systems~\cite{Kugel1982,Feiner1997,Khaliullin1997,Okamoto2001,Maekawa2004,Igoshev2023}.
We note that the definition of the bond-dependent orbital operator $\tau^l$ in Eq.~(\ref{eq:tau}) is different from that in $e_g$ orbital systems, since the rotation of the $\pi_g^*$ orbitals follows the rule in Eq.~(\ref{eq:orbital-rotation}), which is different from that for $e_g$ orbitals.
The KK-type interaction for $\pi$ electron systems has also been derived in the context of organic conductors~\cite{Naka2010,Hotta2010} and \ce{RbO2}~\cite{Ylvisaker2010,Wohlfeld2011}.

Figure~\ref{fig:interaction} shows examples of the bond-dependent orbital interactions in Eq.~(\ref{eq:Heff}).
For the $l=a$ bond, $\tau^l$ is given by $\tau^l=T^z$ and hence the ($\pi_a$, $\pi_b$) basis is relevant.
Depending on the sign of the coefficient for $T^z_i T^z_j$, either $\pi_a$ or $\pi_b$ is uniformly aligned [ferro-orbital (FO) order] or $\pi_a$ and $\pi_b$ orbitals are alternately aligned [antiferro-orbital (AFO) order].
The $l=b$ bond has the same interaction $T^z_i T^z_j$, since $\tau^l=-T^z$.
The difference between $l=a$ and $l=b$ arises in the operator $(1/2+\tau_i^l)$, which projects onto the $\pi_a$ ($\pi_b$) orbital for $l=a$ ($l=b$).
On the other hand, the interaction is described by the $\tau^l=T^x$ ($-T^x$) operator for $l=a+b$ and $l=\textrm{BC}$ ($l=a-b$ and $l=\overline{\textrm{BC}}$). Therefore, if the interactions of the diagonal in the $a$-$b$ plane or the body-center are dominant, the orbital tends to form $\pi_{a+b}$ or $\pi_{a-b}$ orbitals.

\section{Mean-field calculations}

\subsection{Calculation details}

We apply the mean-field (MF) approximation to search for possible phase transitions emerging in the effective model in Eq.~(\ref{eq:Heff}).
The order parameters include $\expval*{T_i^{\xi}}$, $\expval{S_i^z}$, and $\expval*{S_i^z T_i^{\xi}}$, where $\xi=x,y,z$ and $\expval{\cdot}$ stands for the thermal average.
We set $\expval{S_i^x}=\expval{S_i^y}=\expval*{S_i^x T_i^{\xi}}=\expval*{S_i^y T_i^{\xi}}=0$ to enforce the spin moment along the $S^z$ direction, because the spin orientation is arbitrary in the present model without spin-orbit coupling.

\begin{figure}[t]
    \centering
    \includegraphics[width=\linewidth]{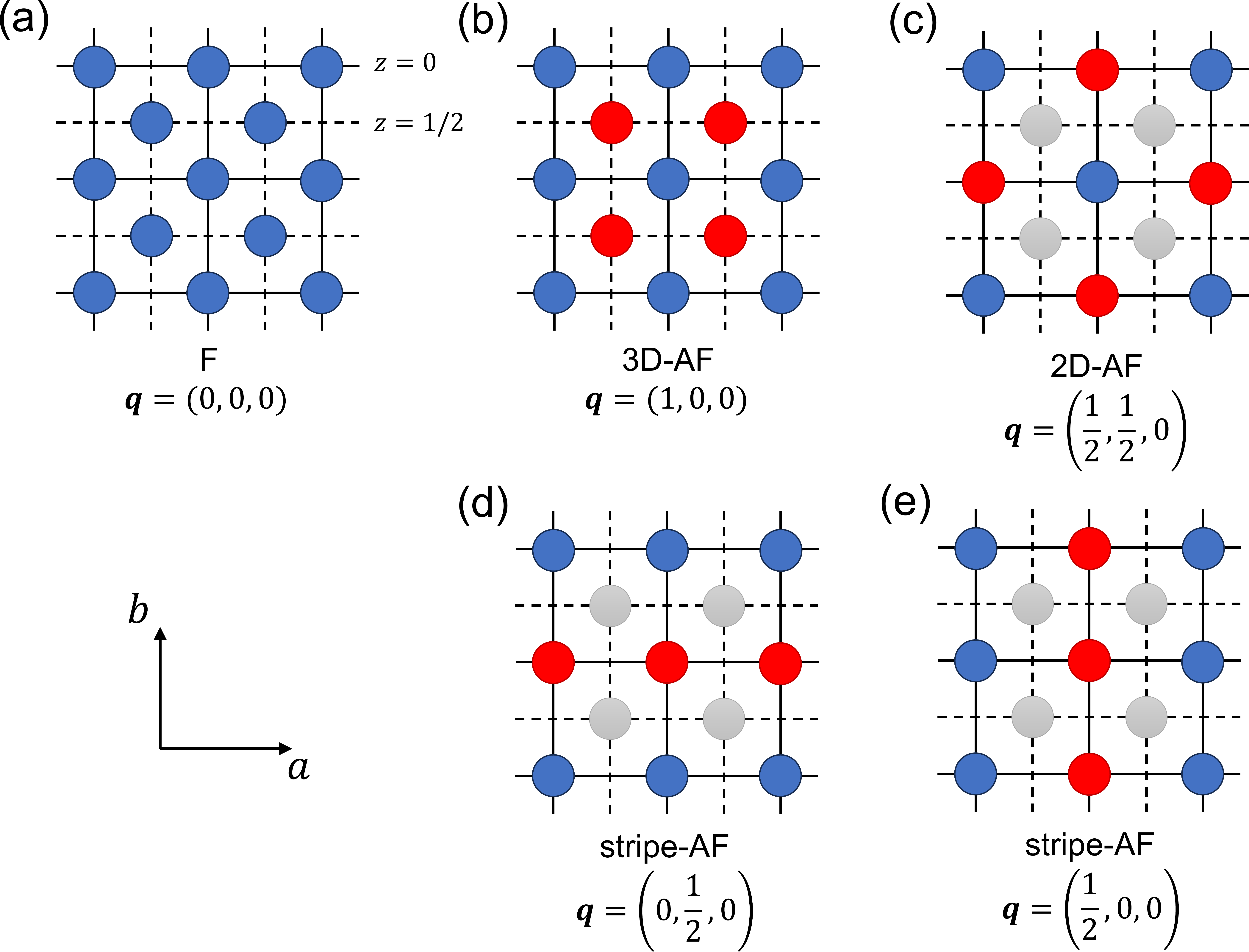}
    \caption{Configurations of the ordered states. The ordering vector $\bm{q}$ is indicated in units of the reciprocal lattice vectors of the conventional unit cell. The circles with the same color represent the same spin or orbital state. The gray sites are not involved in the two-dimensional ordered states in (c)--(e). The sites connected by the solid (dashed) lines represent the layer at $z=0$ ($z=1/2$), respectively. States are labeled F for ferro-orbital/ferromagnetic and AF for antiferro-orbital/antiferromagnetic.}
    \label{fig:order_pattern}
\end{figure}

We consider eight sublattices in a $2\times 2\times 2$ supercell.
Five possible configurations are shown in Fig.~\ref{fig:order_pattern}.
There are (a) one ferroic (F) and (b)--(e) four antiferroic (AF) configurations.
3D-AF (b) is the three-dimensional AF configuration with the ordering vector $\bm{q}=(1, 0, 0)$ [which may also be expressed as $\bm{q}=(0, 1, 0)$ or $\bm{q}=(0, 0, 1)$].
The rest are two-dimensional AF configurations: 2D-AF (c) is the AF order on the square lattice with the ordering vector $\bm{q}=(1/2, 1/2, 0)$. (d) and (e) are stripe-AF with the ordering vector $\bm{q}=(0, 1/2, 0)$ and $\bm{q}=(1/2, 0, 0)$, respectively.
They are degenerate in a tetragonal model.
For reference, various ordered states in the Heisenberg model on the bcc lattice are discussed in Ref.~\cite{Ghosh2019}.

The spin and orbital separately take one of the five configurations in Fig.~\ref{fig:order_pattern}.
We address spin (orbital) configurations by adding M (O) at the end of the configuration name (here, M stands for magnetic or magnetism). Examples include 3D-AFM and stripe-AFO order.

We define the Fourier components of the order parameters as $\expval*{T^{\xi}_{\bm{q}}} = (1/N)\sum_{i} \expval*{T^{\xi}_i} e^{-i\bm{q}\cdot \bm{R}_i}$, where $\bm{R}_i$ is the coordinate of site $i$ and $N$ is the number of sites. 
We use a simplified notation by replacing $\bm{q}$ with the name of the configuration in Fig.~\ref{fig:order_pattern}. For example, $\expval*{T^{\xi}}_\textrm{3D-AF}$ stands for $\expval*{T^{\xi}_{\bm{q}}}$ with $\bm{q}=(1,0,0)$ and $\expval*{S^{z}}_\textrm{stripe-AF}$ stands for $\expval*{S^{z}_{\bm{q}}}$ with $\bm{q}=(1/2,0,0)$ or $(0,1/2,0)$.

There are four interaction parameters, $U$, $U'$, $J_\mathrm{H}$, and $J_\mathrm{H}'$.
We use the standard relations $U'=U-2J_\mathrm{H}$ and $J_\mathrm{H}'=J_\mathrm{H}$ that are valid in $e_g$ orbital systems.
Once the ratio $J_\mathrm{H}/U$ is given, the effective Hamiltonian $\mathcal{H}_\mathrm{eff}$ in Eq.~(\ref{eq:Heff}) is proportional to $J \equiv t^2/U$.
Hence, we vary $J_\mathrm{H}/U$, and measure the temperature $T$ in units of $J$.
As a reference, the values of $U$ and $J_\mathrm{H}$ were estimated for \ce{KO2} using a constrained DFT scheme~\cite{Solovyev2008}, which gives
$U\approx 3.55$\,eV and $J_\mathrm{H}\approx 0.62$\,eV, and thus $J_\mathrm{H}/U \approx 0.17$.
Another estimate on solid oxygen yields $U \approx 11.6$\,eV $J_{\rm H} \approx 0.82$\,eV, and thus $J_{\rm H}/U \approx 0.07$, by the van-der-Waals density functional plus $U$ method~\cite{Kasamatsu2017} and optical absorption experiment~\cite{Landau1962}.
From these estimates, we fix the ratio as $J_{\rm H}/U = 0.1$ in the following calculations, unless otherwise noted.

Regarding the hopping term, there are three parameters $\theta$, $\phi$, and $r^l$.
We fix $r^l$ to the value for {\cso} [(iv) in Table~\ref{tab:hopping}] and vary $\theta$ and $\phi$ to get a comprehensive understanding of the present model.
This will highlight the importance of the geometrical frustration in {\cso} compared to \ce{KO2} and \ce{RbO2}.
Then, we focus on \ce{CsO2} and \ce{RbO2} and discuss the influence of the distortion in the low-temperature phase in Sec.~\ref{sec:ortho}.

\subsection{Ground-state phase diagram for the tetragonal structure}

\begin{figure}[t]
    \centering
    \includegraphics[width=\linewidth]{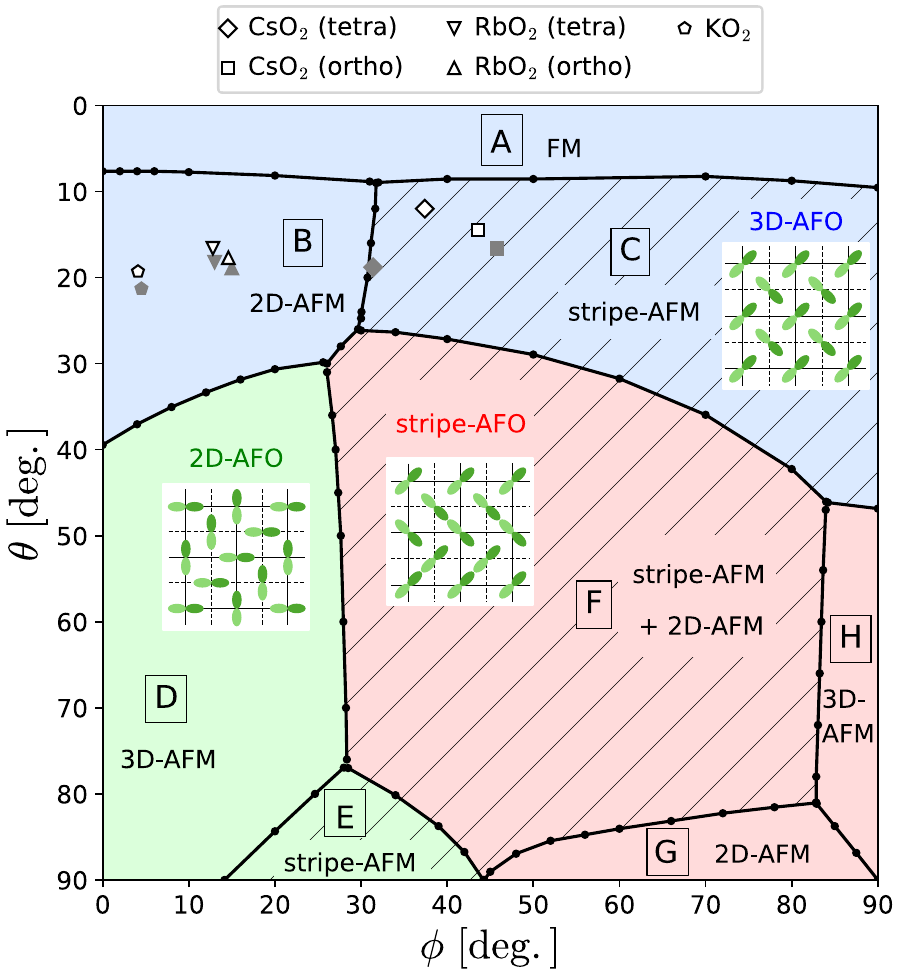}
    \caption{The ground-state phase diagram of the tetragonal model in the $(\phi, \theta)$ plane for $J_\mathrm{H}/U=0.10$.
    The background colors distinguish orbital states. The diagonally shaded areas indicate phases having stripe-AFM order.
    The symbols indicate the DFT estimates for \ce{CsO2}, \ce{RbO2}, and \ce{KO2} (see Table~\ref{tab:hopping}). The open symbols are for the optimized O $z$ positions, and the filled symbols are for the experimental values. The spin-orbital configuration of each phase is shown in Fig.~\ref{fig:ordered_state}. 
    The values of $r^l$ were set to (iv) in Table~\ref{tab:hopping}.}
    \label{fig:phase-diagram}
\end{figure}

\begin{figure*}[tb]
    \centering
    \includegraphics[width=0.8\linewidth]{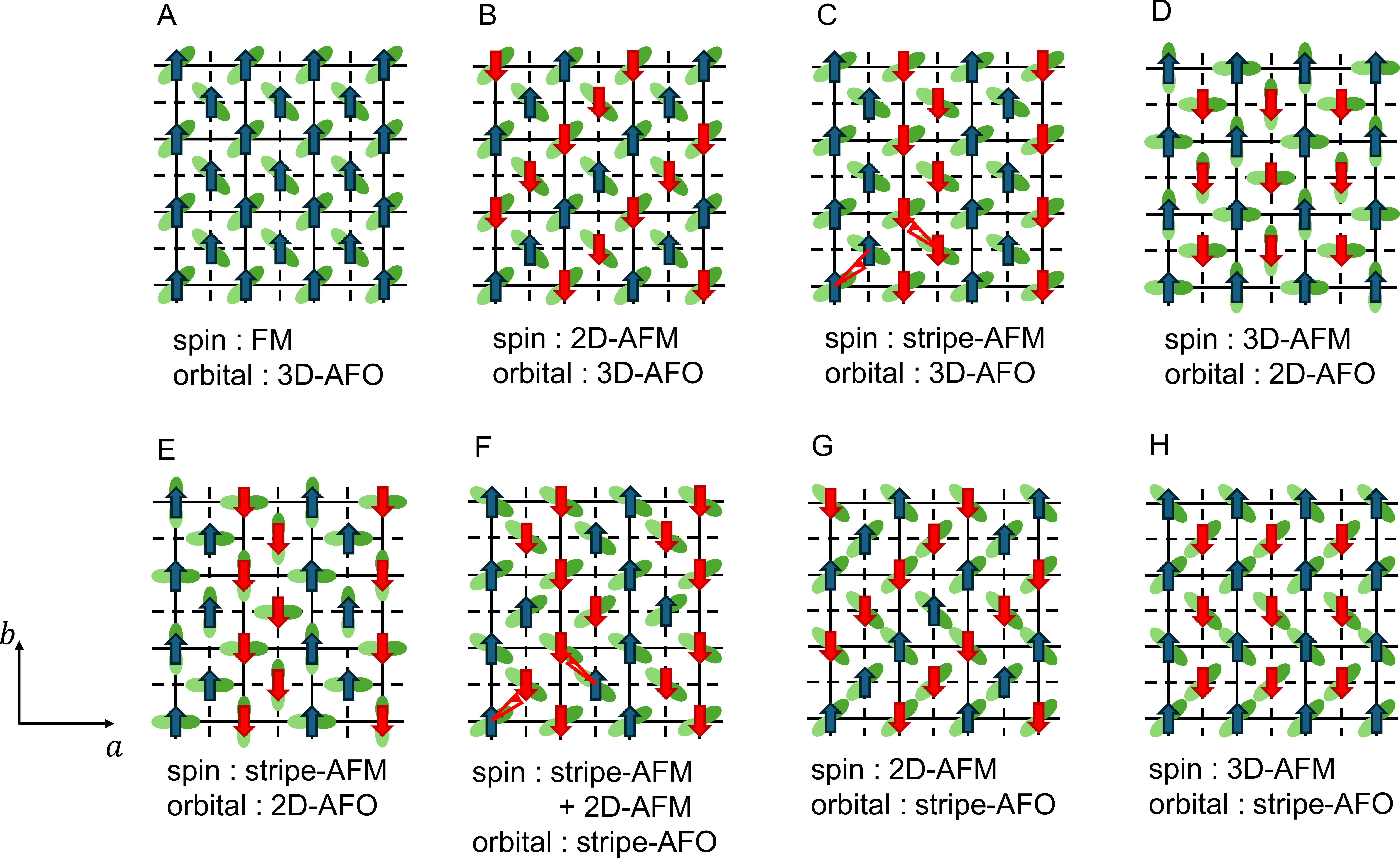}
    \caption{Schematic diagrams of the spin-orbital ordered states appearing in the phase diagram for the tetragonal model in Fig.~\ref{fig:phase-diagram}, The arrows represent spins.
    The stripe-AF orders have two degenerate states with $\bm{q}=(1/2,0,0)$ and $\bm{q}=(0,1/2,0)$. The one stabilized under orthorhombic distortion with $a<b$ is shown.}
    \label{fig:ordered_state}
\end{figure*}

\begin{figure}[t]
    \centering
    \includegraphics[width=\linewidth]{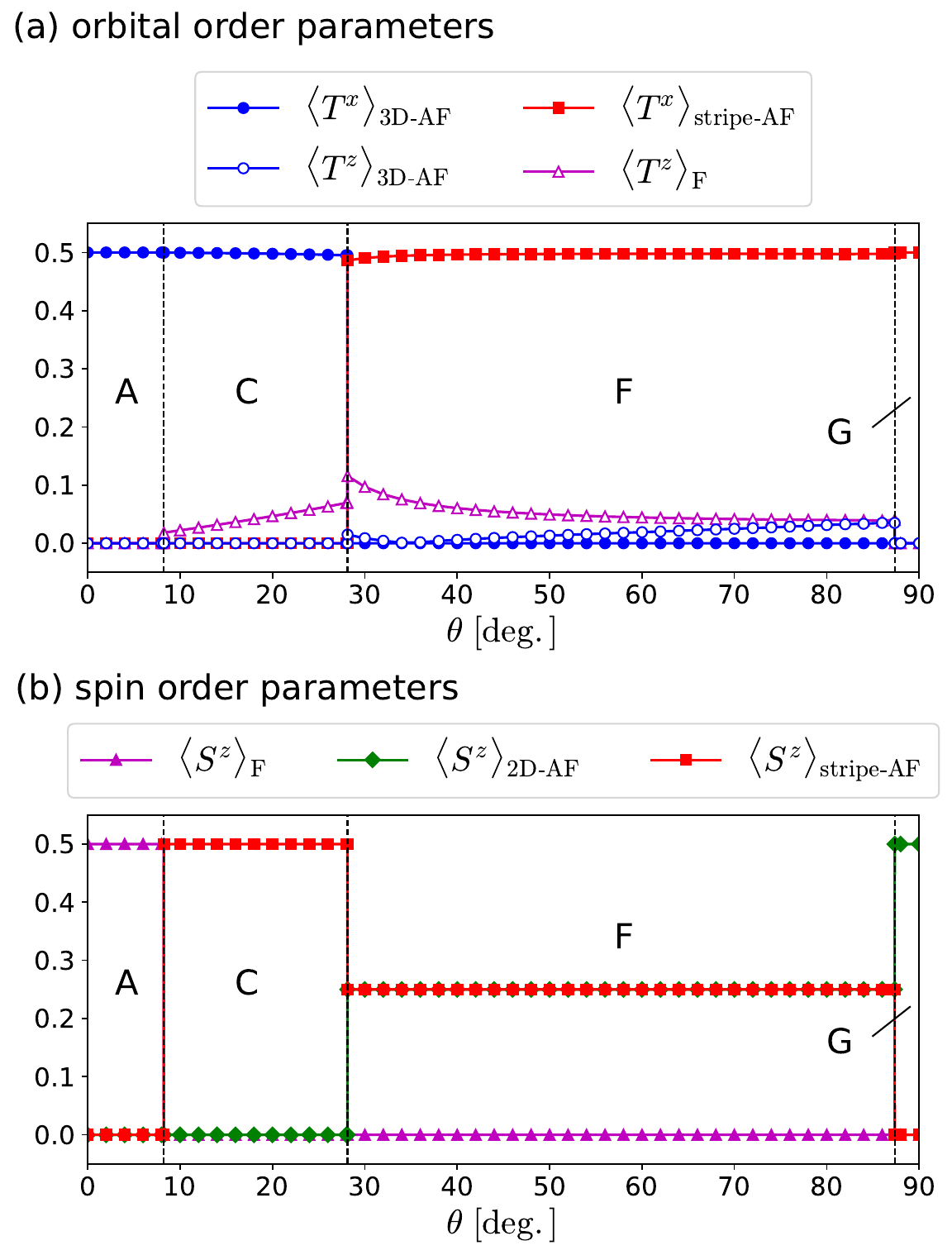}
    \caption{(a) The orbital and (b) spin order parameters in the ground state as functions of $\theta$ for $\phi=46^{\circ}$ and $J_\mathrm{H}/U=0.1$. 
    The labels such as $\expval{S^z}_\textrm{F}$ and $\expval{T^x}_\textrm{3D-AF}$ stand for the Fourier components, where the subscript indicates the configuration in Fig.~\ref{fig:order_pattern}. 
    A, C , F, and G represent the labels of spin-orbital ordered phases listed in Fig.~\ref{fig:ordered_state}.
    The hopping parameters are set to (iv) in Table~\ref{tab:hopping}.}
    \label{fig:theta_order}
\end{figure}

The ground-state phase diagram in the $\phi$-$\theta$ plane is shown in Fig.~\ref{fig:phase-diagram}, where the values of $r^{ab}$, $r^{a+b}$, and $r^{\rm BC}$ are fixed to the parameter set (iv) in Table~\ref{tab:hopping}.
The blue, green, and red regions represent the orbital ordered phases with the propagation vectors $\bm{q}=(1,0,0)$ (3D-AFO order), $\bm{q}=(1/2,1/2,0)$ (2D-AFO order), and $\bm{q}=(0,1/2,0)$ or $(1/2,0,0)$ (stripe-AFO order), respectively. 
The diagonally hatched area corresponds to the stripe-AFM phases, and the other areas are the FM, 2D-AFM, or 3D-AFM ordered phases.

There are eight kinds of ordered phases, termed A-H as shown in Fig.~\ref{fig:ordered_state}, by the combinations of the spin order and orbital patterns and the ordered components of the orbital moment. 
The overall trend is that the 3D-AFO ordered phases are stabilized in the small $\theta$ region ($\theta < 40^\circ$), while in the large-$\theta$ region, the 2D-AFO and stripe-AFO ordered phases compete with each other and a phase transition occurs around $\phi=30$--$40^\circ$. This is because small $\theta$ means that $(t_{\pi}^\textrm{BC})^2$ dominates over $(t_{\pi}^\textrm{ab})^2$ and $(t_{\pi}^{\textrm{a}+\textrm{b}})^2$, and the opposite is true for large $\theta$.
The phase competition between these three orbital ordered states is reproduced by the MF approximation on the orbital-only model obtained by setting the spin operators in Eq.~\eqref{eq:Heff} to zero, presented in Appendix~\ref{sec:orbital_only_model}.
Since the change of the spin configurations strongly depends on the underlying orbital order patterns as well as on $\phi$ and $\theta$, the details will be described in the following.

Here, let us investigate changes of order parameters as a function of $\theta$ while fixing $\phi = 46^\circ$, corresponding to the hopping parameters of tetragonal CsO$_2$.
Figure~\ref{fig:theta_order}\,(a) shows the Fourier components of the orbital order parameters, $\expval{T^x}_\textrm{3D-AF}$, $\expval{T^x}_\textrm{stripe-AF}$, and $\expval{T^z}_\textrm{F}$; $\expval{T^x}_\textrm{3D-AF}$ and $\expval{T^x}_\textrm{stripe-AF}$ represent staggered alignments of $\pi_{a+b}$ and $\pi_{a-b}$ orbitals along the $c$ axis and the $a$ or $b$ axis, respectively, and $\expval{T^z}_\textrm{F}$ denotes the uniform order of $\pi_a$ or $\pi_b$ orbital.
In Fig.~\ref{fig:theta_order}\,(b), the spin order parameters, $\expval{S^z}_\textrm{F}$, $\expval{S^z}_\textrm{stripe-AF}$, and $\expval{S^z}_\textrm{2D-AF}$ are plotted, which characterize the FM, stripe-AFM, and 2D-AFM order of the $S^z$ component, respectively.
All the order parameters are defined so that their maximum values are $0.5$.

At $\theta = 0$, $\expval{T^x}_\textrm{3D-AF}$ and $\expval{S^z}_\textrm{F}$ take the value $0.5$, showing that the phase A (FM + 3D-AFO order) shown in Fig.~\ref{fig:ordered_state} is realized.
When $\theta$ increases, $\expval{T^x}_\textrm{3D-AF}$ slightly decreases from $0.5$ and instead $\expval{T^z}_\textrm{F}$ becomes finite and increases above $\theta \simeq 8^\circ$.
Simultaneously, the $\expval{S^z}_\textrm{F}$ discontinuously drops to zero and  $\expval{S^z}_\textrm{stripe-AF}$ jumps to the maximum value.
This is a first-order phase transition from phase A to C (stripe-AFM + 3D-AFO order).
We note that in phase C the direction of the $\pi$ orbital shows canting towards the $a$ axis due to the small $\expval{T^z}_\textrm{F}$ as shown in Fig.~\ref{fig:ordered_state}.

Further increasing $\theta$ in Fig.~\ref{fig:theta_order}, the dominant orbital order parameter changes to $\expval{T^x}_\textrm{stripe-AF}$ and the canting component $\expval{T^z}_\textrm{F}$ begins to decrease for $\theta \gtrsim 28^\circ$.
As for the spin sector, $\expval{S^z}_\textrm{stripe-AF}$ discontinuously decreases and coexists with $\expval{S^z}_\textrm{2D-AF}$.
This is because the stripe-AFM and 2D-AFM orders separately develop on the $z=0$ and $1/2$ planes, respectively, as the phase F in Fig.~\ref{fig:ordered_state}. 
These results indicate that the ground state changes from phase C to phase F at $\theta \simeq 28^\circ$.
When $\theta$ approaches the maximum value $90^\circ$, only two of the constituents in phase F, $\expval{T^x}_\textrm{stripe-AF}$ and $\expval{S^z}_\textrm{2D-AF}$, remain finite and the canting of the orbitals vanishes (phase G).

For $\phi \lesssim 44^\circ$ in the phase diagram in Fig.~\ref{fig:phase-diagram}, phases B, D, and E (see Fig.~\ref{fig:ordered_state}) appear in the region with $\theta > 10^\circ$.
Phase B has the same orbital configuration (3D-AFO order of $T^x$) with phase A, while the spin pattern is 2D-AFM.
Increasing $\theta$ from phase B, the system enters phase D, where the orbital and spin configurations change to the 2D-AFO order of $T^z$ and 3F-AFM order, respectively.
In phase E, which is stabilized for $\theta \gtrsim 80^\circ$, the orbital configuration is the same as in the neighboring phase D whereas the spin configuration is stripe-AFM, which is common to the other neighboring phase F.
On the other hand, in phase H near $\phi = 90^\circ$, the 3D-AFM order appears on the stripe-AFO order, which is shared with phases F and G.

The DFT estimates of $(\theta, \phi)$ for \ce{CsO2}, \ce{RbO2}, and \ce{KO2} (Table~\ref{tab:hopping}) are indicated by symbols in Fig.~\ref{fig:phase-diagram}, showing that \ce{CsO2} is located in phase C but in the competing region with adjacent phases A, B, and F.
We note that, since other parameters $r^l=t_{\delta}^l/t_{\pi}^l$ were fixed at the value for \ce{CsO2}, the symbols of \ce{RbO2} and \ce{KO2} should be take just for reference; however, as we will show below, the full set of parameters still leads to phase B for \ce{RbO2}.

So far we have fixed the value of $J_\mathrm{H}/U$ at 0.1. If $J_\mathrm{H}/U$ is increased with $(\theta, \phi)$ fixed at the value for \ce{CsO2}, phases A and B are stabilized, while phase F is stabilized if $J_\mathrm{H}/U$ is decreased. We obtain phase C in the range $0.03 \lesssim J_\mathrm{H}/U \lesssim 0.13$. See Appendix~\ref{sec:JH_dependence} for details.

\subsection{Influence of distortion}
\label{sec:ortho}

In the previous subsection, we have investigated stable spin and orbital ordered states in the tetragonal structure common to the \ce{$A$O2} at room temperature. 
Here, we discuss the influence of the lattice distortion which depends on each compound at low temperatures, especially focusing on the orthorhombic \ce{CsO2} and monoclinic \ce{RbO2}. 
Since \ce{KO2} shows a triclinic structure accompanied by tilting of O$_2$ molecules, which is beyond our effective model assuming the O-O bond parallel to the $c$ axis, we leave it for future work.

\subsubsection{Orthorhombic distortion in \ce{CsO2}}

\begin{figure}[t]
    \centering
    \includegraphics[width=\linewidth]{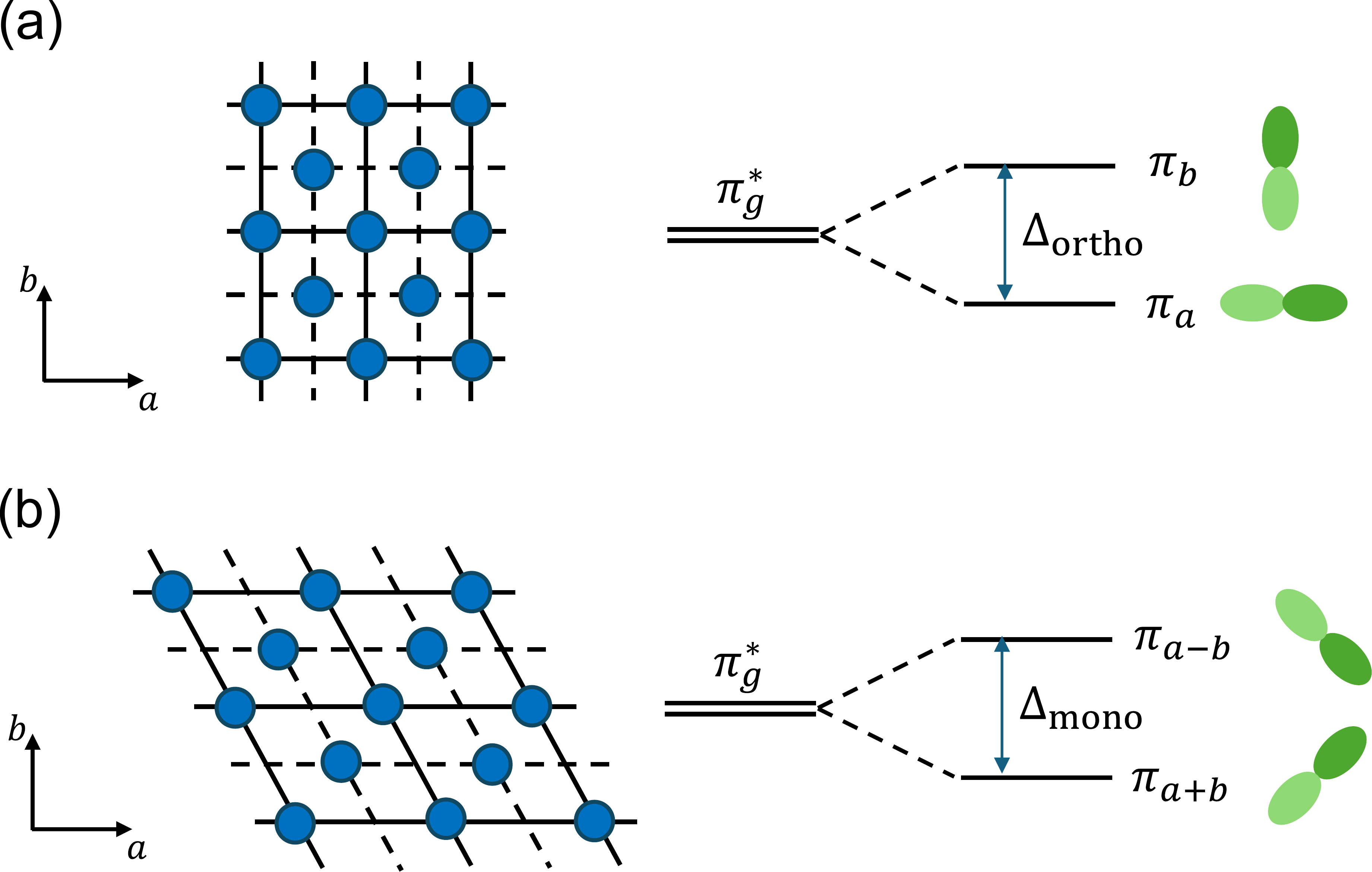}
    \caption{Schematic diagrams of (a) the orthorhombic distortion with $a<b$ and (b) the monoclinic distortion with $\gamma>90^{\circ}$, and the resultant CEF splitting of the $\pi_g^{\ast}$ orbitals.}
    \label{fig:CEF}
\end{figure}

\ce{CsO2} has the orthorhombic structure with $a<b$ for $T<T_\mathrm{s1}\simeq 150$\,K (Fig.~\ref{fig:phases}). 
The orthorhombic crystalline electric field (CEF) lifts the degeneracy between $\pi_a$ and $\pi_b$ orbitals as illustrated in Fig.~\ref{fig:CEF}\,(a).
We note that the distortion $a<b$ stabilizes the $\pi_a$ orbital in the hole picture.
This energy splitting due to the orthorhombic distortion can be represented using the operator $T^z$ as
\begin{align}
    \mathcal{H}_\mathrm{CEF} = -\Delta_\mathrm{ortho} \sum_i T_{i}^z.
    \label{eq:CEF_ortho}
\end{align}
We thus consider the orthorhombic model given by $\widetilde{\mathcal{H}}_\mathrm{eff} = \mathcal{H}_\mathrm{eff} + \mathcal{H}_\mathrm{CEF}$ and adopt the parameter set (ii) in Table.~\ref{tab:hopping} as the hopping parameters for the low-temperature structure of CsO$_2$.

\begin{figure}[t]
    \centering
    \includegraphics[width=\linewidth]{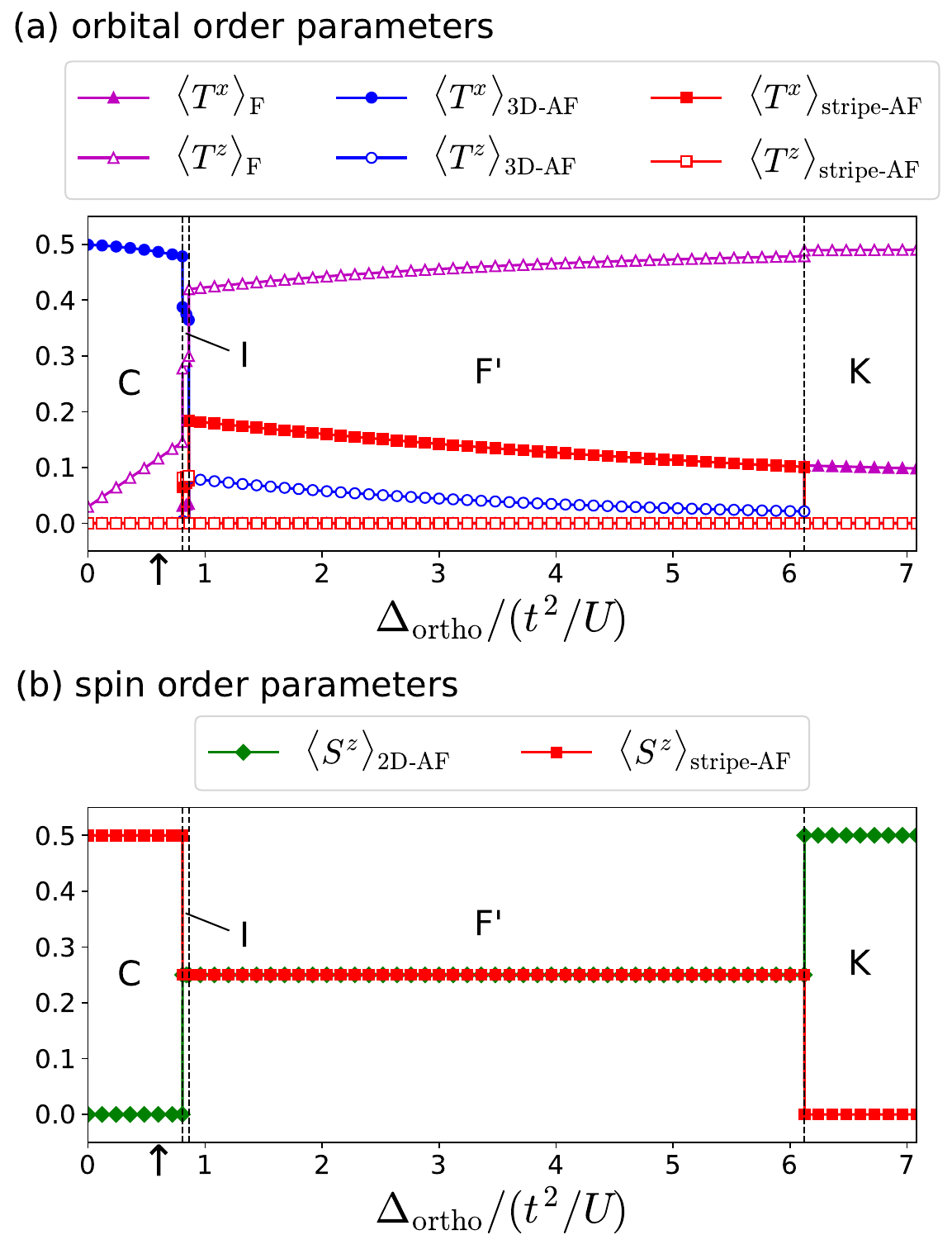}
    \includegraphics[width=\linewidth]{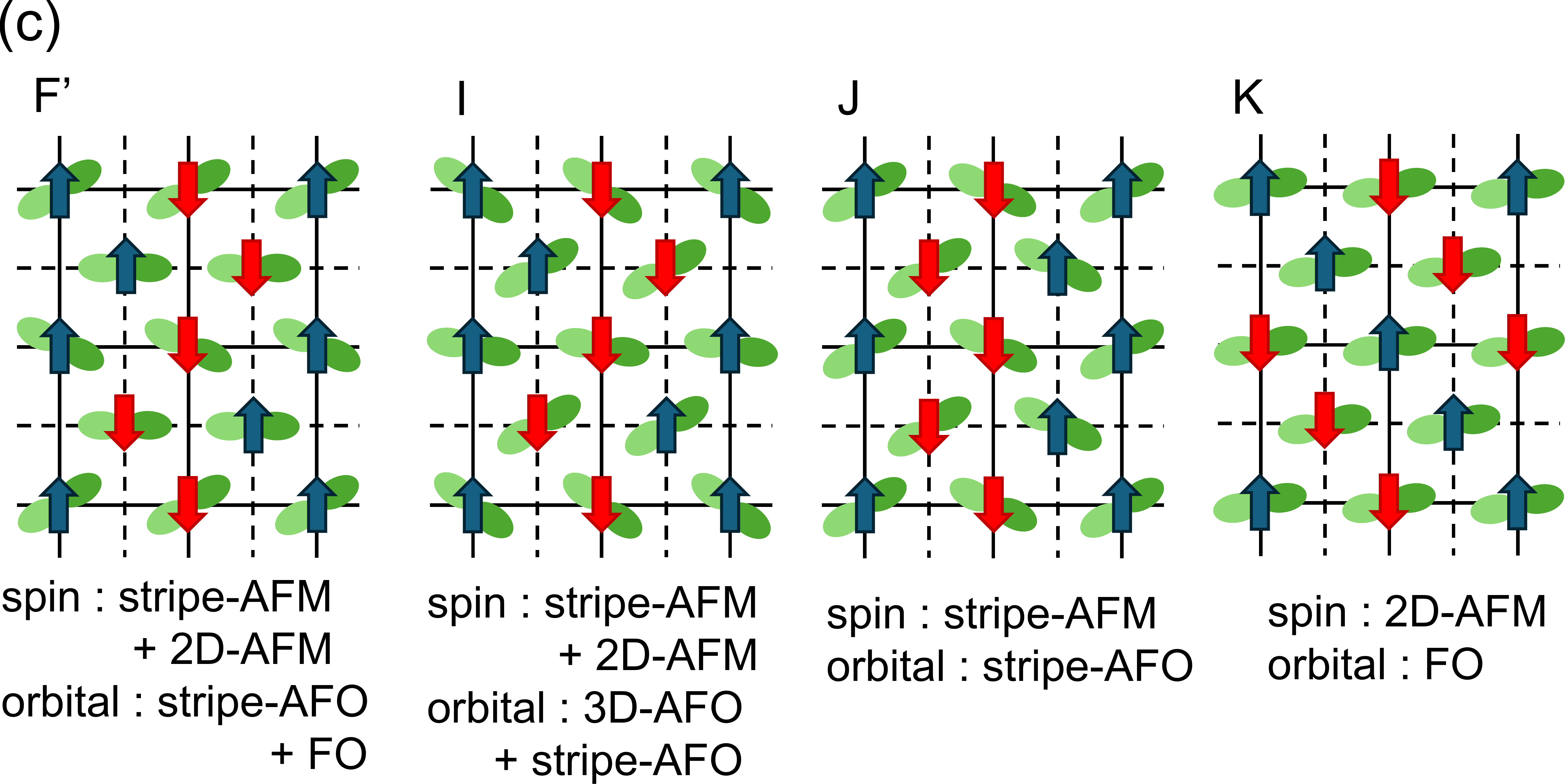}
    \caption{(a) The orbital and (b) spin-order parameters in the orthorhombic model as a function of $\Delta_\mathrm{ortho}$ defined in Eq.~\eqref{eq:CEF_ortho}.
    The hopping parameters~(ii) in Table~\ref{tab:hopping} were used. 
    The arrow represents the DFT estimate for the orthorhombic \ce{CsO2}, $\Delta_\mathrm{ortho}/(t^2/U)=0.60$.
    (c) Schematic ordering patterns in the orthorhombic model.}
    \label{fig:CEF_ortho}
\end{figure}

Figure~\ref{fig:CEF_ortho}\,(a) and \ref{fig:CEF_ortho}\,(b) show the $\Delta_\mathrm{ortho}$ dependence of the orbital and spin order parameters, respectively.
In the absence of $\Delta_\mathrm{ortho}$, the ground state is phase C, which is the same as in the tetragonal CsO$_2$ [parameter set (iv)] shown in Fig.~\ref{fig:phase-diagram}.
When $\Delta_\mathrm{ortho}$ is introduced, $\expval{T^z}_\textrm{F}$, which directly couples to the CEF, monotonically increases and the spin and orbital patterns successively change.
At $\Delta_\mathrm{ortho}/(t^2/U) \simeq 0.81$, $\expval{T^x}_\textrm{3D-AF}$ discontinuously decreases, and simultaneously, other five orbital-order parameters plotted in Fig.~\ref{fig:CEF_ortho}\,(a) become finite.
In the spin sector in Fig.~\ref{fig:CEF_ortho}\,(b), $\expval{S^z}_\textrm{2D-AF}$ and $\expval{S^z}_\textrm{stripe-AF}$ are finite at the same value, indicating the coexistence of the 2D-AFM order and the stripe-AFM order as in phase F.
This ordered state, termed phase I, is represented by the diagram in Fig.~\ref{fig:CEF_ortho}\,(c). Based on the state in phase F, the orbitals are canted to the $a$ axis accompanied by the stripe-AFO order with $\bm{q}=(0, 1/2, 0)$ on the $z=0$ plane.
When $\Delta_\mathrm{ortho}$ is slightly increased in phase I, the ground state changes to phase F$'$, which is stable in a wide range of $\Delta_\mathrm{ortho}$.
The orbital pattern of phase F$'$ is that the $z=1/2$ plane (stripe-AFO order) in phase F is replaced by the FO order of $\pi_a$ orbitals.
Further increasing $\Delta_\mathrm{ortho}$, the orbital pattern of the $z=0$ plane is also forced to FO order, and accordingly, phase K with the 2D-AFM order in all planes, shown in Fig.~\ref{fig:CEF_ortho}\,(c), becomes stable.

\begin{figure}[t]
    \centering
    \includegraphics[width=\linewidth]{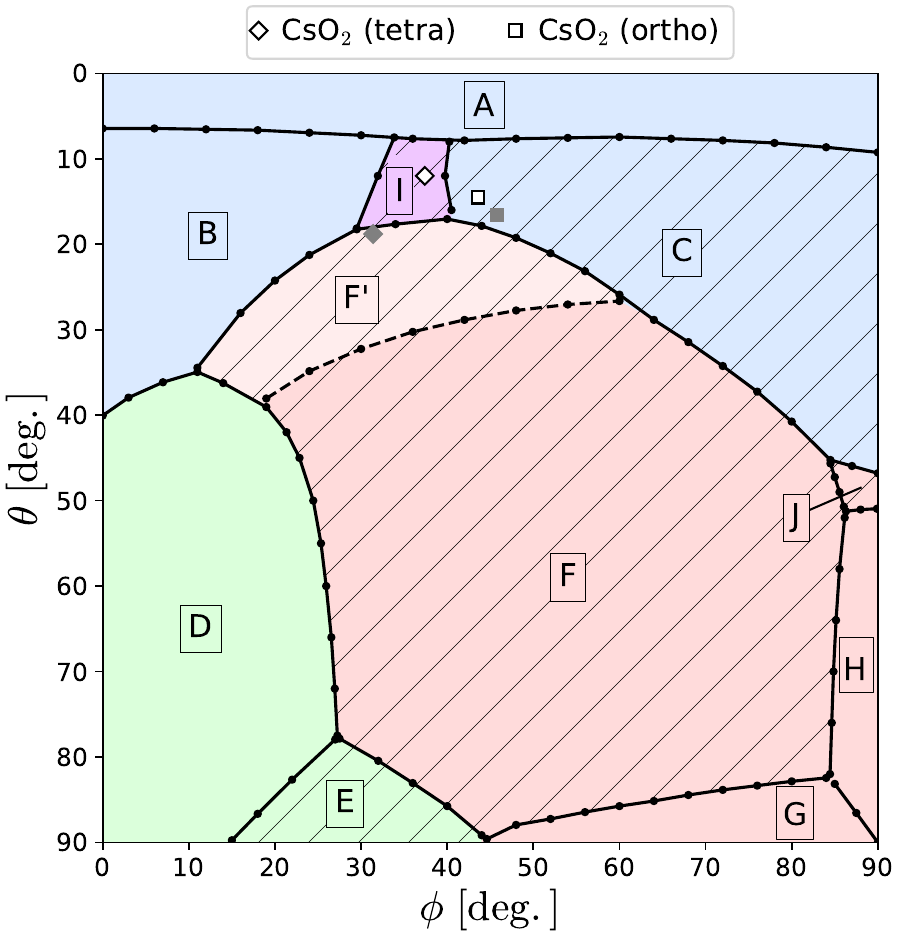}
    \caption{The ground-state phase diagram of the orthorhombic model in the $(\phi, \theta)$ plane.
    The symbols indicate the DFT estimates for \ce{CsO2}.
    See the caption of Fig.~\ref{fig:phase-diagram} for more details. 
    The parameter set (ii) in Table~\ref{tab:hopping} were used with $\Delta_\mathrm{ortho}/(t^2/U)=0.60$.}
    \label{fig:phase-diagram-ortho}
\end{figure}

According to the DFT calculation for orthorhombic CsO$_2$ [parameter set (ii) in Table.~\ref{tab:hopping}], the CEF splitting is estimated to be $\Delta_\mathrm{ortho} = 1.1$ meV, which is normalized as $\Delta_\mathrm{ortho} / (t^2/U) \approx 0.60$, using $t=81$ meV and $U=3.55$ eV~\cite{Solovyev2008}.
This is located in phase C as indicated by arrows in Figs.~\ref{fig:CEF_ortho}\,(a) and \ref{fig:CEF_ortho}\,(b).
Figure~\ref{fig:phase-diagram-ortho} shows the ground-state phase diagram obtained by changing $\phi$ and $\theta$ in the presence of the orthorhombic distortion with $\Delta_\mathrm{ortho} / (t^2/U) = 0.60$.
Comparing with the phase diagram for the tetragonal model in Fig.~\ref{fig:phase-diagram}, we can see that, although the minor phases I and F$'$ develop in the parameter region surrounded by phases A, B, C, D, and F, the overall structure of the phase diagram does not change significantly.
In addition, the narrow phase J appears between phases C and H in the large $\phi$ region, where both the orbital and spin orders are stripe-AF configuration.

\subsubsection{Monoclinic distortion in \ce{RbO2}}

\ce{RbO2} undergoes two structural phase transitions from tetragonal to orthorhombic, $a\neq b$, and then to monoclinic, $\gamma \neq 90^{\circ}$, with decreasing temperature, as shown in Fig.~\ref{fig:phases}.
The distortion of the angle with $\gamma>90^{\circ}$ lifts the degeneracy of the $\pi_g^{\ast}$ orbitals into $\pi_{a+b}$ and $\pi_{a-b}$ orbitals as illustrated in Fig.~\ref{fig:CEF}\,(a).
This energy splitting can be represented using the operator $T^x$. Hence, the CEF potential in the monoclinic phase is given by the combination of $T^x$ and $T^z$ as
\begin{align}
    \mathcal{H}_\mathrm{CEF} = -\sum_i ( \Delta_\mathrm{ortho} T_{i}^z +\Delta_\mathrm{mono} T_{i}^x).
    \label{eq:CEF_monocli}
\end{align}
We estimated $\Delta_\mathrm{ortho}$ and $\Delta_\mathrm{mono}$ by the DFT calculation for the monoclinic structure (not shown), and obtained $\Delta_\mathrm{ortho}=1.15$\,meV and $\Delta_\mathrm{mono}=4.95$\,meV.
We fix the ratio $\Delta_\mathrm{ortho}/\Delta_\mathrm{mono}=0.232$ and vary $\Delta_\mathrm{mono}$ to discuss the influence of the monoclinic CEF in \ce{RbO2}.
The hopping parameter ratios $r^l$ for the monoclinic structure differ from those for the orthorhombic structure only by 0.01. Hence, we adopt the values in (vi) of Table~\ref{tab:hopping}.

\begin{figure}[t]
    \centering
    \includegraphics[width=\linewidth]{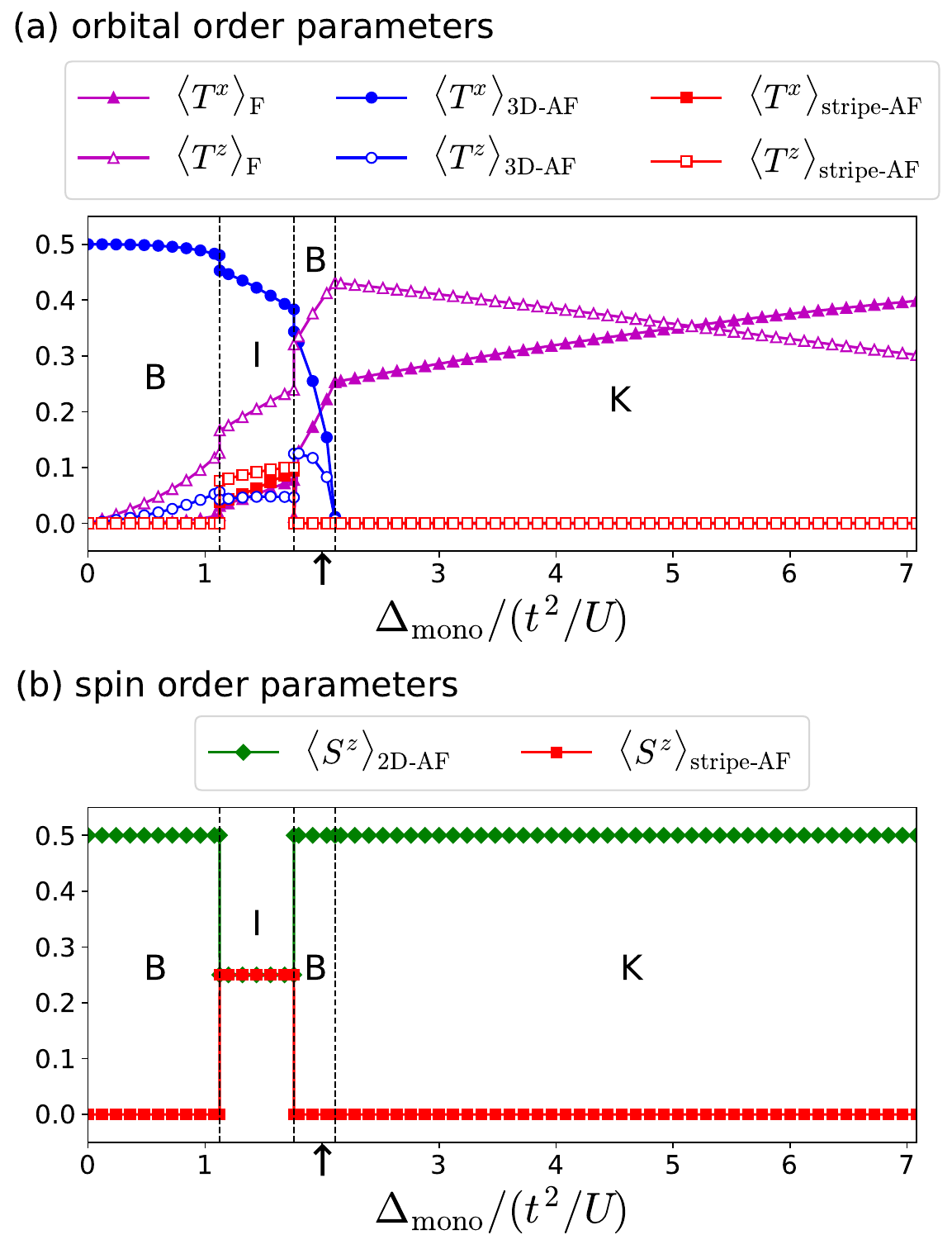}
    \includegraphics[width=\linewidth]{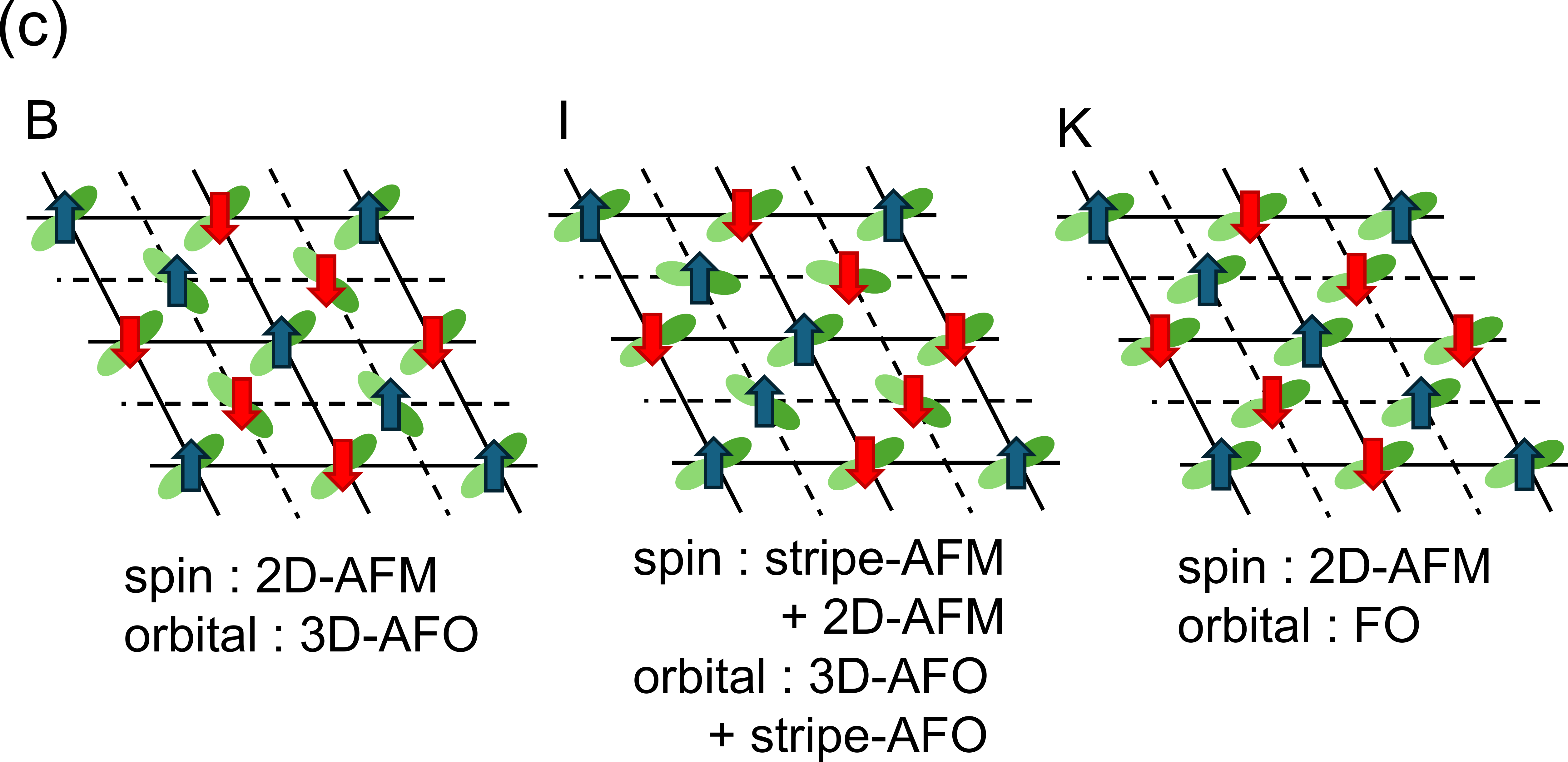}
    \caption{(a) The orbital and (b) spin order parameters in the monoclinic model as a function of $\Delta_\mathrm{mono}$ with fixed $\Delta_\mathrm{ortho}/\Delta_\mathrm{mono}=0.232$. The hopping parameter set (vi) in Table~\ref{tab:hopping} was used.
    The arrow represents the DFT estimate for monoclinic \ce{RbO2}, $\Delta_\mathrm{mono}/(t^2/U)=2.0$.
    (c) Schematic ordering patterns in the monoclinic model.}
    \label{fig:CEF_monocli}
\end{figure}

Figures~\ref{fig:CEF_monocli}\,(a) and \ref{fig:CEF_monocli}\,(b) show the variations of the orbital and spin order parameters as a function of $\Delta_\mathrm{mono}$.
The ordered states in the presence of monoclinic distortion are depicted in Fig.~\ref{fig:CEF_monocli}\,(c).
At $\Delta_\mathrm{mono}=0$, the ground state is phase B (3D-AFO + 2D-AFM order) in contrast to \ce{CsO2}.
In the middle region of $1.13 \leq \Delta_\mathrm{mono} \leq 1.76$ in phase B, six orbital order parameters plotted in Fig.~\ref{fig:CEF_monocli}\,(a) and two spin order parameters in Fig.~\ref{fig:CEF_monocli}\,(b) appear in a first-order transition. This state corresponds to phase I, which appeared also in the case of the orthorhombic distortion.
Further increasing $\Delta_\mathrm{mono}$, the forced FO ordered state with 2D-AFM order is stabilized (phase K).
The difference from the orthorhombic case is that $\expval{T^z}_\textrm{F}$ begins to decrease in the large $\Delta_\mathrm{mono}$ region since the distortion is coupled with $\expval{T^x}_\textrm{F}$.
Besides, under the monoclinic distortion, phase F$'$ shown in Fig.~\ref{fig:CEF_ortho}\,(c) does not appear. This is because $\expval{T^z}_\textrm{F}$ in the $z=1/2$ plane is unstable in the monoclinic CEF in Eq.~\eqref{eq:CEF_monocli}.

The DFT estimate $\Delta_\mathrm{mono}=4.95$\,meV is normalized to $\Delta_\mathrm{mono}/(t^2/U) = 2.0$ using $t=95$\,meV and $U=3.55$\,eV~\cite{Solovyev2008}. This value is indicated by arrows in Figs.~\ref{fig:CEF_monocli}\,(a) and \ref{fig:CEF_monocli}\,(b). We thus conclude that \ce{RbO2} is in phase B or phase K. In both cases, the spin structure is 2D-AFM, which is characterized by the translation vector $\bm{q}=(1/2, 1/2, 0)$.

\subsection{Finite temperature properties}

\begin{figure}[t]
    \centering
    \includegraphics[width=\linewidth]{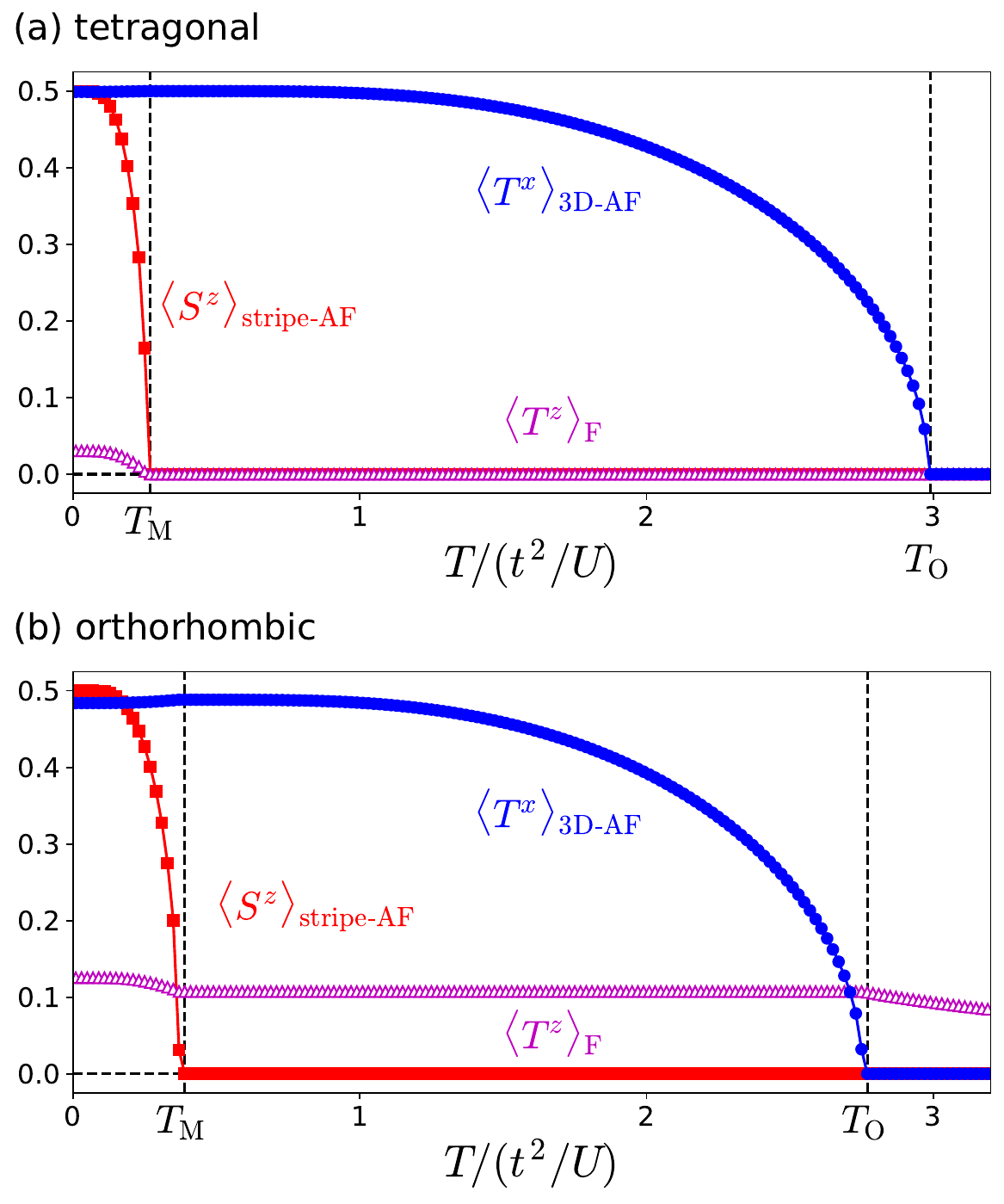}
    \caption{Temperature dependence of the order parameters in phase C. (a) The tetragonal parameter set (iv) in Table~\ref{tab:hopping} and (b) the orthorhombic parameters (ii) with $\Delta_\mathrm{ortho}/(t^2/U)=0.60$ were used.}
    \label{fig:m_T}
\end{figure}

We conclude this section by presenting finite temperature properties.
Figure~\ref{fig:m_T} shows the temperature dependence of the order parameters in phase C.
(a) is the result for the tetragonal model and (b) for the orthorhombic model with finite $\Delta_\mathrm{ortho}$.
The order parameter for phase C is represented by $\expval{T^x}_\textrm{3D-AF}$ for the orbital part and $\expval{S^z}_\textrm{stripe-AF}$ for the spin part.
We define the temperatures of the orbital and spin orders by $T_\mathrm{O}$ and $T_\mathrm{M}$, respectively.
$T_\mathrm{O}$ is about six times higher than $T_\mathrm{M}$. This ratio depends on parameters such as $\theta$, $\phi$, and $J_\mathrm{H}/U$.

It is interesting that the canting of the orbital, represented by $\expval{T^z}_\mathrm{F}$, appears only below $T=T_\mathrm{M}$ in Fig.~\ref{fig:m_T}\,(a). This means that the stripe-AFM order gives rise to the canting of the orbital. In fact, the direction of the stripe-AFM order and the canting of the orbital are correlated with each other.
In the orthorhombic model in Fig.~\ref{fig:m_T}\,(b), $\expval{T^z}_\mathrm{F}$ is finite in the whole temperature range because of the external orthorhombic distortion.
$\expval{T^z}_\mathrm{F}$ exhibits a cusp at $T=T_\mathrm{O}$ and increases below $T=T_\mathrm{M}$, indicating that the stripe-AFM order \textit{enhances} the orthorhombic distortion.

\section{Discussion}

\subsection{Role of the orbital degree of freedom}

In this section, we first discuss the role of the orbital degree of freedom in our results.
In particular, we focus on phase C (stripe-AFM + 3D-AFO order), which corresponds to the orthorhombic phase of CsO$_2$, and consider the origin of the phase transitions.
The exchange interactions on all three kinds of bonds are relevant.
The order of their strengths is $|J^\mathrm{BC}| > |J^{a}| \simeq |J^{a+b}|$ as shown in Fig.~\ref{fig:polar}.
The leading interaction $J^\mathrm{BC}$ leads to the 3D-AFO ordered state as demonstrated in the orbital-only model in Appendix~\ref{sec:orbital_only_model}.
The spin correlations are then considered on top of the 3D-AFO ordered state.

\begin{figure}[t]
    \centering
    \includegraphics[width=0.45\linewidth]{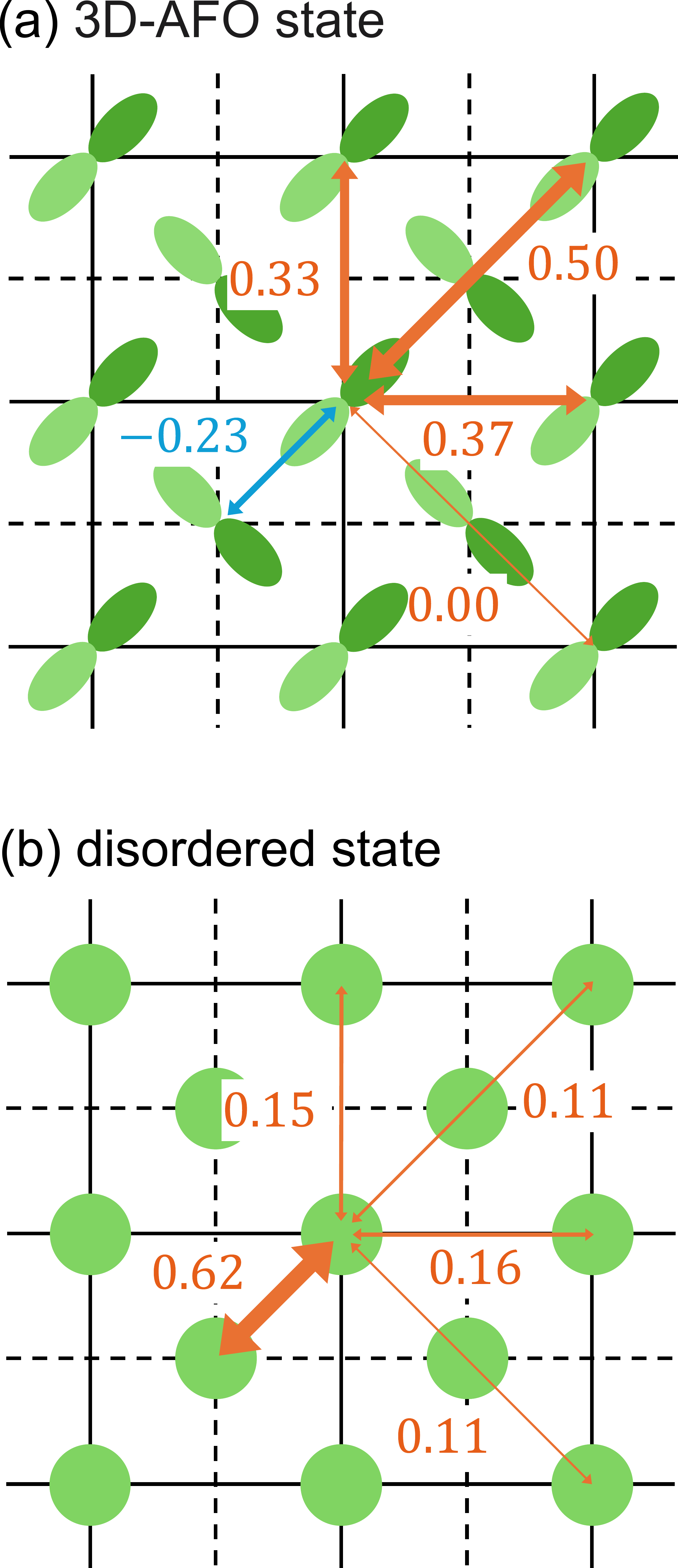}
    \caption{Values of the coupling constants $J_{ij}$ in units of $t^2/U$ in the effective Heisenberg model in Eq.~\eqref{eq:Heisenberg}. The orange and blue indicate AFM and FM interactions, respectively.
    (a) 3D-AFO ordered state, (b) disordered state.
    The parameter set for the orthorhombic \ce{CsO2} in (ii) of Table~\ref{tab:hopping} was used.}
    \label{fig:Heisenberg}
\end{figure}

For this purpose, we derive the effective spin-spin interactions represented by the Heisenberg Hamiltonian
\begin{align}
    \mathcal{H}_\mathrm{spin} = \sum_{\langle ij \rangle} J_{ij} \bm{S}_i \cdot \bm{S}_j,
    \label{eq:Heisenberg}
\end{align}
by eliminating the orbital operators from $\mathcal{H}_\mathrm{eff}$ in Eq.~\eqref{eq:Heff}.
We estimate $J_{ij}$ by replacing the orbital operators $T_i^{z}$ and $T_i^{x}$ with their expectation values. In the 3D-AFO ordered state in phase C, for example, $T_i^{x}$ is replaced by $+1/2$ or $-1/2$ depending on the site and $T_i^{z}$ is replaced by 0 for all sites (the canting of the orbital is ignored). Figure~\ref{fig:Heisenberg}\,(a) shows the exchange interactions $J_{ij}$ obtained in the 3D-AFO ordered state.
The leading interaction turns out to be the AFM interaction on the diagonal bond in the $a$-$b$ plane ($l=a+b$),  which favors the stripe-AFM ordering~\cite{Jin2013,Glasbrenner2015}.
For comparison, we estimated $J_{ij}$ in the disordered state by replacing all orbital operators with zero in $\mathcal{H}_\mathrm{eff}$.
The result is presented in Fig.~\ref{fig:Heisenberg}\,(b). The strengths of $J_{ij}$ in the disordered state are simply determined by the hopping amplitude. Therefore, the $l=\textrm{BC}$ bond has the largest AFM interaction, which does not enhance the stripe-AFM order. The comparison between Figs.~\ref{fig:Heisenberg}\,(a) and (b) clearly demonstrates that the orbital order in the 3D-AFO ordered state is relevant for the emergence of the stripe-AFM state. Sensitivity of the magnetic order to the presence or absence of orbital order has been observed for \ce{RbO2}~\cite{Nandy2010}.

Finally, we consider the direction of the stripe-AFM state under the orthorhombic distortion.
The external orthorhombic distortion with $a<b$ tilts the $\pi_{a+b}$ and $\pi_{a-b}$ orbitals to the $a$ axis (we again note that we are considering holes). 
Therefore, the interaction on the $a$ bond becomes predominant over the $b$ bond since the $a$ ($b$) bond is described more by $\pi$ ($\delta$) hopping under the distortion.
The KK mechanism explains the AF spin configuration on the ferro-orbital configuration on the $a$ bond, which leads to the stripe-AFM order with the translation vector $\bm{q}=(1/2,0,0)$.

\subsection{Implication for experiments}

\subsubsection{\ce{CsO2}}

Recent neutron scattering experiments for \ce{CsO2} reported the stripe-AFM order with propagation vector $\bm{q}=(0, 1/2, 0)$~\cite{Nakano2023} or $\bm{q}=(1/2, 0, 0)$~\cite{Ewings2023} in the orthorhombic structure with $a<b$.
Our results for phase C and other phases having the stripe-AFM ordered configuration exhibit $\bm{q}=(1/2, 0, 0)$ because the KK interaction on the $l=a$ bond favors $\bm{q}=(1/2, 0, 0)$ over $\bm{q}=(0, 1/2, 0)$ as discussed above.

The DFT estimate is located in phase C but close to phases A, B, I, and F$'$ in the phase diagram in Fig.~\ref{fig:phase-diagram-ortho}. Among those nearby phases, I and F$'$ have the stripe-AFM configuration. However, it is not a pure stripe-AFM order but a stack of the stripe-AFM and the 2D-AFM orders.
In these phases, neutron scattering experiments should observe not only $\bm{q}=(1/2, 0, 0)$ but also $\bm{q}=(1/2, 1/2, 0)$. Since the peak at $\bm{q}=(1/2, 1/2, 0)$ has not been observed, we exclude phases I and F$'$ and propose only phase C as a candidate for \ce{CsO2}.

\begin{figure}[t]
    \centering
    \includegraphics[width=\linewidth]{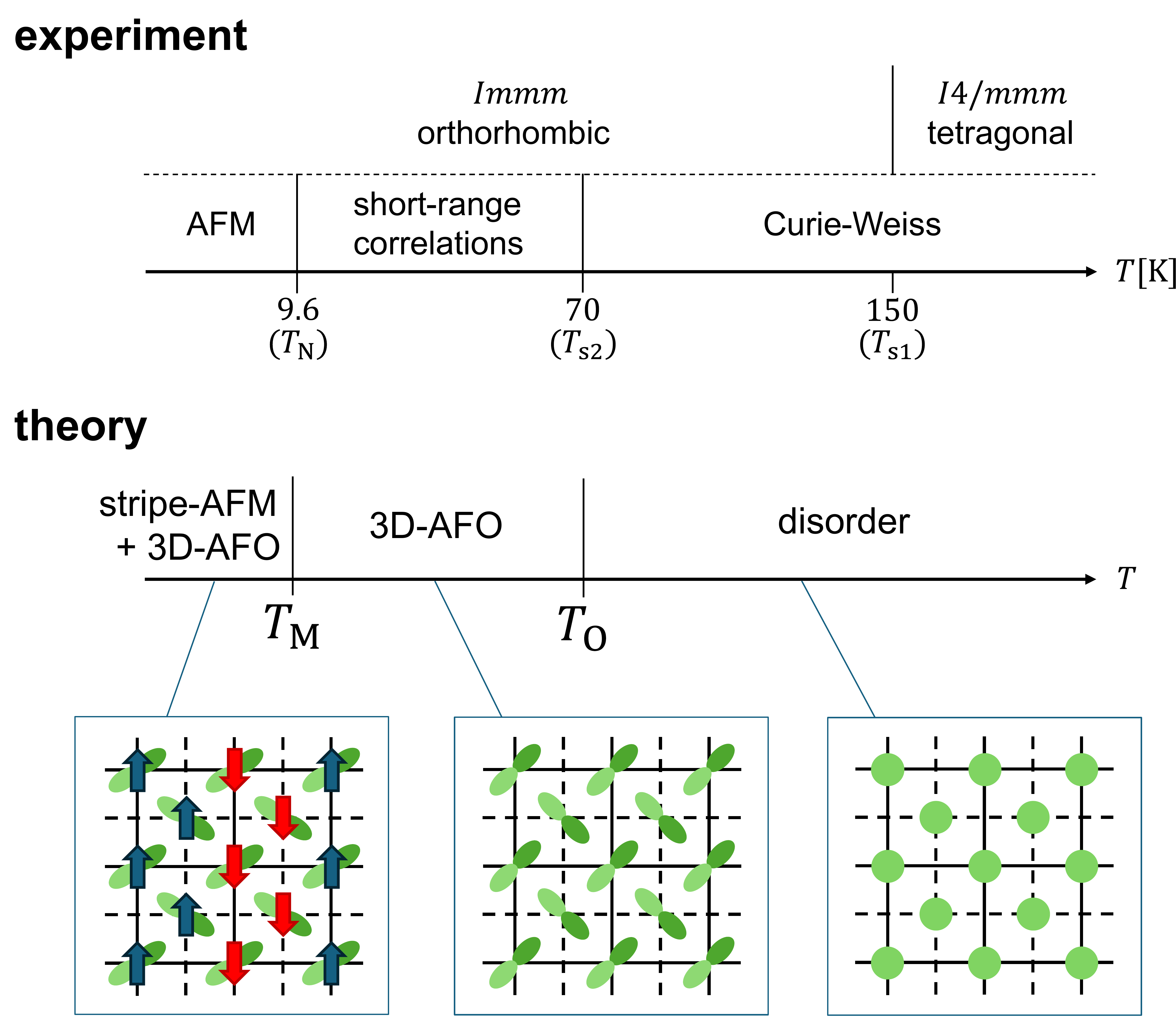}
    \caption{Comparison between the experimental and theoretical finite-$T$ phase diagrams.}
    \label{fig:scenario}
\end{figure}

We turn our attention to the finite-temperature properties in {\cso}.
Experimentally, there are three phase transitions, $T_\mathrm{s1}$, $T_\mathrm{s2}$, and $T_\mathrm{N}$, in phases II--III (Fig.~\ref{fig:phases}).
On the other hand, we obtained two phase transitions in phase C of our model: the stripe-AFM transition at $T_\mathrm{M}$ and the 3D-AFO order transition at $T_\mathrm{O}$ [Fig.~\ref{fig:m_T}\,(a) for tetragonal structure and Fig.~\ref{fig:m_T}\,(b) for orthorhombic structure].
The energy unit $J\equiv t^2/U$ is estimated to be $J \approx 21$\,K using the DFT value $t=81$\,meV for the orthorhombic {\cso} in (ii) of Table~\ref{tab:hopping} and $U=3.55$\,eV~\cite{Solovyev2008}. Hence, $T_\mathrm{M}$ and $T_\mathrm{O}$ in Fig.~\ref{fig:m_T}\,(b) are converted to $T_\mathrm{M} \approx 8.4$\,K and $T_\mathrm{O} \approx 59$\,K.
We identify $T_\mathrm{M}$ with $T_\mathrm{N}$ since the calculated magnetic structure and the transition temperature is consistent with the experiment.
We further identify $T_\mathrm{O}$ with $T_\mathrm{s2}$ as presented in Fig.~\ref{fig:scenario} since the AFO ordered state for $T_\mathrm{M} < T < T_\mathrm{O}$ does not give rise to a global lattice distortion as shown in Fig.~\ref{fig:m_T}\,(a).
The experimental structural phase transition from tetragonal to orthorhombic at $T=T_\mathrm{s1}$ is ascribed to an origin that is not considered in our effective model, e.g., the Jahn-Teller effect. Describing this transition would require a model that also includes lattice degrees of freedom.

In this scenario, the magnetic properties observed for $T_\mathrm{N}<T<T_\mathrm{s2}$ should be explained by the 3D-AFO ordered state.
Experimentally, the temperature dependence of the susceptibility for $T_\mathrm{N}<T<T_\mathrm{s2}$ is well fitted by the Bonner-Fisher function~\cite{Riyadi2012,Miyajima2018}, which was developed to fit the one-dimensional Heisenberg model. In our model with \ce{O2}--\ce{O2} hopping, none of the ordered states in Fig.~\ref{fig:ordered_state} gives a one-dimensional hopping path.
We note that even the stripe-AFO ordered state proposed in Ref.~\cite{Riyadi2012} (e.g., $z=0$ plane in phase F) is not one-dimensional in the presence of the \ce{O2}--\ce{O2} direct hopping.
Considering the fact that the Bonner-Fisher curve of the susceptibility can be observed in a wide range of Heisenberg models~\cite{Zheng2005}, 
we expect that the susceptibility could be reproduced by the frustrated Heisenberg model on top of the 3D-AFO order as presented in Fig.~\ref{fig:Heisenberg}\,(a).
For this, strong correlations and thermal fluctuations beyond the MF approximation need to be included which is beyond the scope of the present study.

The neutron diffraction experiment in Ref.~\cite{Ewings2023} also suggests doubling of the unit cell along the $a$ axis for $T \leq T_{\rm s2}$, 
which is attributed to displacements of Cs and O$_2$ ions along the $b$ axis with the propagation vector $\bm{q}=(1/2, 0, 0)$.
On the other hand, in the present calculation, the stripe-AFO order with $\bm{q}=(1/2, 0, 0)$ has not been obtained within the realistic parameter range. 
Therefore, this structural change is expected to originate from the instability of the background lattice system rather than the correlated $\pi$-electron system.

Finally, we refer to the experimental indication for the stripe-AFM + 3D-AFO ordered state of phase C in CsO$_2$.
As shown in Fig.~\ref{fig:m_T}, the orbital order parameter $\expval{T^z}_\mathrm{F}$ corresponding to the orthorhombic distortion is enhanced below $T_{\rm M}$, accompanied by the development of the stripe-AFM order.
This is because in phase C the orbital on each site tends to cant uniformly towards the $a$ axis
to gain the AFM exchange interaction along the $a$ axis, parallel to the spin propagation vector.
This canting of the orbital moment can be detected as the elongation and contraction of the $a$ and $b$ axes, respectively, as the temperature decreases through $T_{\rm N}$.
This prediction provides a good validation of our scenario in experiments.

\subsubsection{\ce{RbO2} and \ce{KO2}}

Regarding \ce{RbO2}, the spin structure has not been resolved experimentally (apart from a study of oxygen deficient \ce{RbO2}~\cite{Riyadi2011}). Our results in Fig.~\ref{fig:CEF_monocli} predict phase B, namely, the 2D-AFM order with $\bm{q}=(1/2, 1/2, 0)$ on top of the 3D-AFO order, which is the same orbital order as in \ce{CsO2}. In \ce{KO2}, magnetic order occurs in the monoclinic phase which is beyond the scope of the present study. A previous theoretical study~\cite{Wohlfeld2011} discussed the 3D-AFO + 2D-AFM order (phase D in Fig.~\ref{fig:phase-diagram}) for \ce{KO2} and \ce{RbO2}. It is close to our DFT estimates, but in-plane hopping would need to be stronger for its realization.

\section{Summary}

We investigated the spin-orbital order in \ce{$A$O2} ($A=\textrm{Cs}$, Rb, K) using a strong-coupling effective model derived based on first-principles calculations.
Relevant interactions between the $\pi_g^{\ast}$ orbitals on the \ce{O2} molecule are up to third neighbor for tetragonal and up to fourth neighbor for orthorhombic structures. It is common to all $A$ atoms that the interaction between the corner and body-center sites ($l=\textrm{BC}$) is the largest.
The difference in $A$ atoms can be seen in the $a$-$b$ plane. {\cso} has highly frustrated interactions between the $a$ ($b$) bond and the diagonal $a+b$ bond, while \ce{RbO2} and \ce{KO2} have weaker frustration.
We conclude that a relevant microscopic control parameter that distinguishes the low-temperature properties in \ce{$A$O2} is the magnitude of the geometrical frustration in the $a$-$b$ plane ($\phi$ in our notation).

The MF calculations for the strong-coupling effective model reveal possible ground states in {\cso}. Based on this, we propose a 3D-AFO order of $\pi_{a+b}$ and $\pi_{a-b}$ orbitals with the ordering vector $\bm{q}=(1, 0, 0)$ below $T=T_\mathrm{s2} \simeq 70$\,K. This orbital order is caused by the leading interaction on the BC bond. The subleading interactions (those in the $a$-$b$ plane) determine the spin structure. The stripe-AFM state with $\bm{q}=(1/2, 0, 0)$ is realized due to the frustrated interactions in the $a$-$b$ plane. We predict that the canting of the orbital moment towards the $a$ axis, which is consistent with the orthorhombic distortion with $a<b$, is enhanced, i.e., the lattice distortion increases, by the stripe-AFM order below $T_\mathrm{N}=9.6$\,K.
For \ce{RbO2}, we predict the 2D-AFM order with $\bm{q} = (1/2, 1/2, 0)$ due to weaker frustration.

The peculiar magnetic properties below $T=T_\mathrm{s2}$ in {\cso} remain unsolved in the present MF calculations. Our results suggest that the temperature dependence of the susceptibility fitted by the Bonner-Fisher curve should be explained by another mechanism such as geometrical frustration in systems with two or three dimensions. The fact that the observed magnetic moment is considerably reduced from $S=1/2$~\cite{Nakano2023,Ewings2023} indicates the importance of correlations in the magnetic properties, which is left for future work.

\begin{acknowledgments}
We acknowledge useful discussions with T. Kambe, T. Nakano, and J. Nasu.
This work was supported by JSPS KAKENHI Grants No. 20K20522, No. 21H01003, No. 21H01041, No. 23H04869, No. 19K03723, No. 23H01129, and No. 23K25826.
\end{acknowledgments}

\appendix

\section{Coupling constants}
\label{sec:Js}

\begin{figure}[t]
    \centering
    \includegraphics[width=\linewidth]{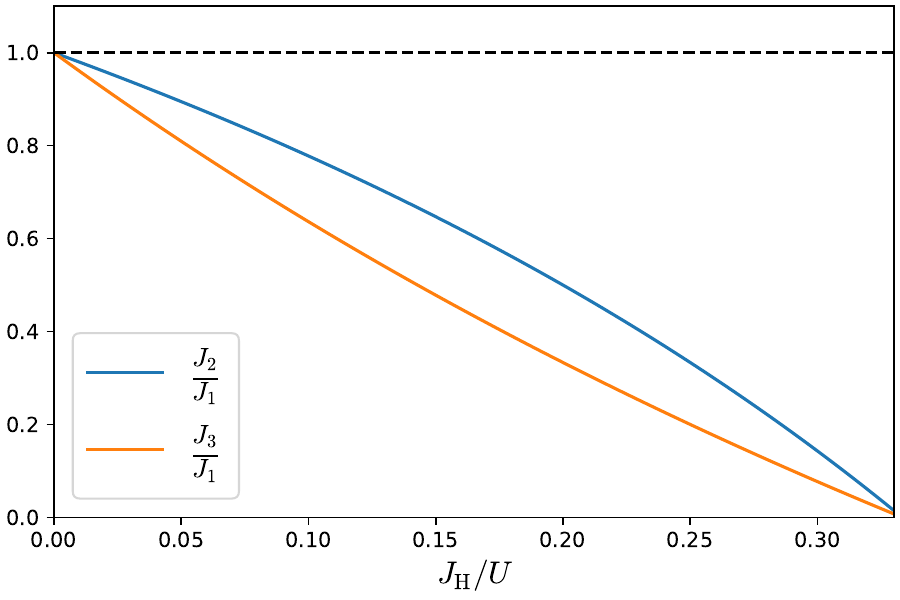}
    \caption{The ratios $J_2^l/J_1^l$ and $J_3^l/J_1^l$ as a function of $J_\mathrm{H}/U$.}
    \label{fig:Js}
\end{figure}

In this appendix, we compare the magnitude of the coupling constants defined in Eq.~(\ref{eq:Js}).
The ratio between $J_1^l$, $J_2^l$, $J_2^{'l}$, and $J_3^l$ is determined by the local interaction parameters.
Figure~\ref{fig:Js} shows the ratios, $J_2^l/J_1^l$ and $J_3^l/J_1^l$ as a function of $J_\mathrm{H}/U$. Here, we use the relation $U'=U-2J_\mathrm{H}$ and $J_\mathrm{H}'=J_\mathrm{H}$ as in the main text. In this case, $J_2^{'l}$ is identical to $J_2^l$, namely, $J_2^{'l}=J_2^l$.
Figure~\ref{fig:Js} demonstrates the inequality $J_1^l \geq J_2^l = J_2^{'l} \geq J_3^l$. The equality holds when $J_\mathrm{H}/U=0$.
For $J_\mathrm{H}/U=0.1$, for example, $J_2^l$ and $J_3^l$ are about 80{\%} and 60{\%} of $J_1^l$, respectively.

\section{Orbital-only model}
\label{sec:orbital_only_model}

\begin{figure}[t]
    \centering
    \includegraphics[width=0.9\linewidth]{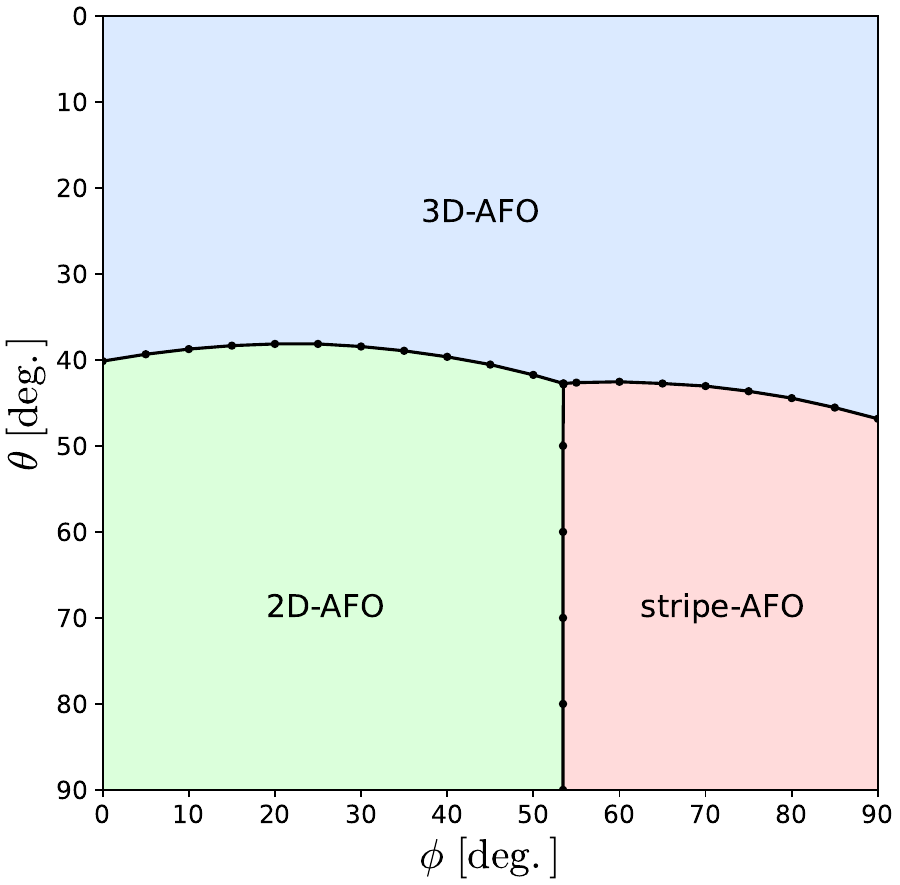}
    \includegraphics[width=\linewidth]{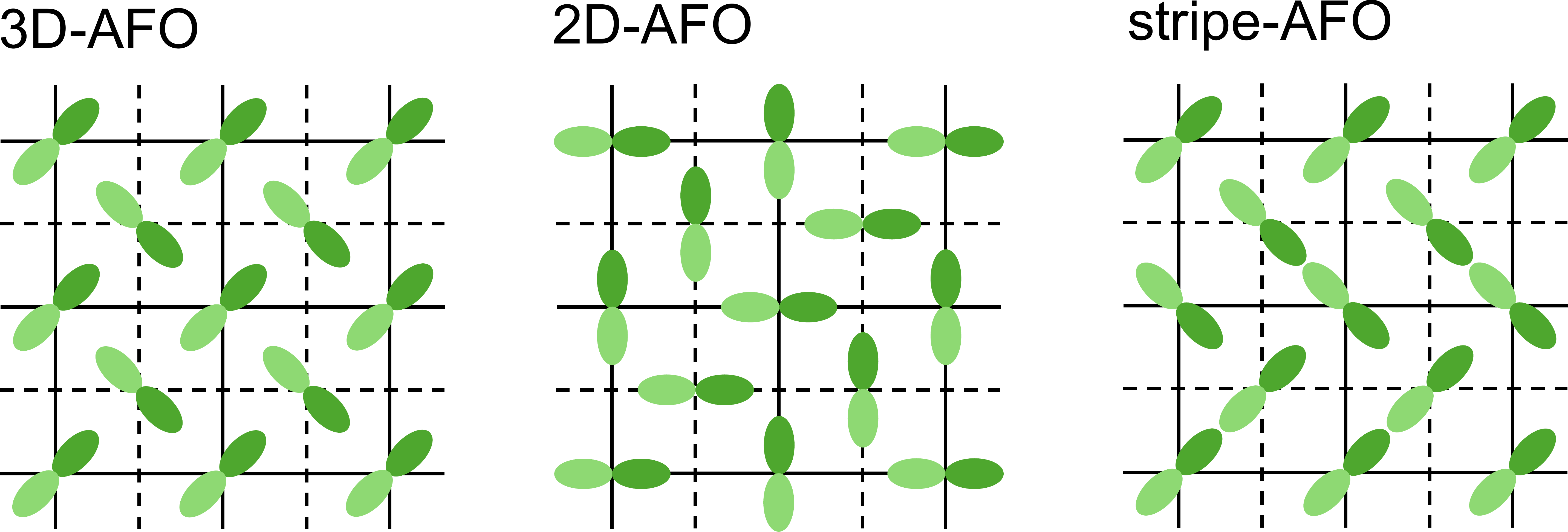}
    \caption{The ground-state phase diagram and orbital configuration in the orbital-only model. The parameters are the same as in Fig.~\ref{fig:phase-diagram}.}
    \label{fig:pure_orbital}
\end{figure}

The effective Hamiltonian $\mathcal{H}_\mathrm{eff}$ in Eq.~(\ref{eq:Heff}) consists of the spin and orbital operators. Here, we consider an orbital-only model by setting $\bm{S}_i=0$ in $\mathcal{H}_\mathrm{eff}$.
Figure~\ref{fig:pure_orbital} shows the ground-state phase diagram of the orbital-only model in the $(\phi, \theta)$ plane.
This figure explains the tendency of the orbital order in Fig.~\ref{fig:phase-diagram}.

There are three phases. 
The 3D-AFO state of the $(\pi_{a+b}, \pi_{a-b})$ orbital is stabilized due to the interaction between the corner site and the body-center site, which is dominant near $\theta=0$. 
The region near $\theta=90^{\circ}$ is divided into two phases with a two-dimensional character.
The 2D-AFO state of the $(\pi_{a}, \pi_{b})$ orbital is stabilized below $\phi \simeq 53^{\circ}$ by the interaction along the $a$ axis and $b$ axis.
The diagonal interaction in the $a$-$b$ plane, which is dominant near $\phi=90^{\circ}$, stabilizes the stripe-AFO order of the $(\pi_{a+b}, \pi_{a-b})$ orbital.

\section{$J_\mathrm{H}/U$ dependence}
\label{sec:JH_dependence}
\begin{figure}[t]
    \centering
    \includegraphics[width=\linewidth]{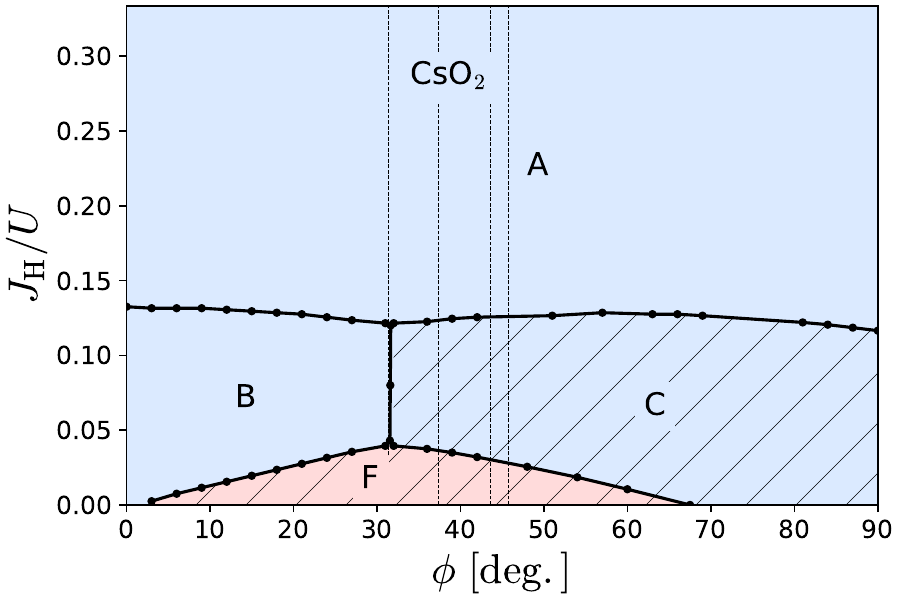}
    \caption{
    The ground-state phase diagram in $(\phi, J_\mathrm{H}/U)$ plane with $\theta=12.0^{\circ}$. The parameter set for the tetragonal \ce{CsO2} was used [(iv) in Table~\ref{tab:hopping}].
    The vertical dashed lines indicate the DFT estimates of the $\phi$ value for \ce{CsO2} (see Table~\ref{tab:hopping}).}
    \label{fig:beta_J}
\end{figure}

In the main text, $J_\mathrm{H}/U$ has been fixed at 0.1. In this appendix, we present how the phases change as $J_\mathrm{H}/U$ is varied. 
Figure~\ref{fig:beta_J} shows the phase diagram with $\phi$ and $J_\mathrm{H}/U$ on the axes. The value of $\theta$ is fixed at $\theta=12^{\circ}$ for the tetragonal {\cso}.
The cut of Fig.~\ref{fig:beta_J} at $J_\mathrm{H}/U=0.1$ corresponds to the horizontal cut of Fig.~\ref{fig:phase-diagram} at $\theta=12^{\circ}$.
It turns out that $J_\mathrm{H}/U$ stabilizes phase A over phases B and C, whereas phase F appears when $J_\mathrm{H}/U$ is decreased.
Comparison between Fig.~\ref{fig:beta_J} and Fig.~\ref{fig:phase-diagram} indicates that an increase of $J_\mathrm{H}/U$ corresponds to a decrease of $\theta$.
This tendency can be understood as follows. As $J_\mathrm{H}/U$ increases, the $J_1^l$ term becomes dominant (Appendix~\ref{sec:Js}). Then, the simple KK-type ordered state is favored. In contrast, four interaction terms ($J_1^l$, $J_2^l$, $J_2^{'l}$, and $J_3^l$) become relevant in the limit $J_\mathrm{H}/U \to 0$. Competition between different interaction terms favors rather complicated states such as phase F.

\bibliography{refs}

\begin{thebibliography}{48}
\bibitem [{I4m()}]{I4mmm_note}%
  \BibitemOpen
  \href@noop {} {}\bibinfo {note} {$I4/mmm$ high symmetry points are
  $X=(0,0,\nicefrac{1}{2})$, $Y=(-\zeta,\zeta,\nicefrac{1}{2})$,
  $\Sigma=(-\eta,\eta,\eta)$,
  $Z=(\nicefrac{1}{2},\nicefrac{1}{2},-\nicefrac{1}{2})$,
  $\Sigma_1=(\eta,1-\eta,-\eta)$, $N=(0,\nicefrac{1}{2},0)$,
  $P=(\nicefrac{1}{4},\nicefrac{1}{4},\nicefrac{1}{4})$,
  $Y_1=(\nicefrac{1}{2},\nicefrac{1}{2},-\zeta)$ where
  $\eta=\frac{1}{4}\big(1+\frac{a^2}{c^2}\big)$ and
  $\zeta=\frac{a^2}{2c^2}$.}\BibitemShut {Stop}%
\bibitem [{Imm()}]{Immm_note}%
  \BibitemOpen
  \href@noop {} {}\bibinfo {note} {$Immm$ high symmetry points are
  $X=(-\zeta,\zeta,\zeta)$, $L=(-\mu,\mu,\nicefrac{1}{2}-\delta)$,
  $T=(0,0,\nicefrac{1}{2})$,
  $W=(\nicefrac{1}{4},\nicefrac{1}{4},\nicefrac{1}{4})$,
  $R=(0,\nicefrac{1}{2},0)$, $X_1=(\zeta,1-\zeta,-\zeta)$,
  $Z=(\nicefrac{1}{2},\nicefrac{1}{2},-\nicefrac{1}{2})$,
  $Y=(\eta,-\eta,\eta)$, $S=(\nicefrac{1}{2},0,0)$ where
  $\zeta=\frac{1}{4}\big(1+\frac{a^2}{c^2}\big)$,
  $\eta=\frac{1}{4}\big(1+\frac{b^2}{c^2}\big)$, $\delta=\frac{b^2-a^2}{4c^2}$
  and $\mu=\frac{a^2+b^2}{4c^2}$.}\BibitemShut {Stop}%
\end{thebibliography}

\end{document}